\RequirePackage{fix-cm}
\documentclass[twocolumn]{svjour3}          
\smartqed  
\usepackage{graphicx}
\usepackage{epsfig,epstopdf}

\usepackage[pagebackref=true,breaklinks=true,letterpaper=true,colorlinks,bookmarks=false]{hyperref}

\usepackage{hyperref}
\usepackage{url}
\usepackage{adjustbox}
\usepackage{multirow}
\usepackage{caption}
\usepackage{subcaption}
\usepackage{epsfig,epstopdf}
\usepackage{graphicx}
\usepackage{amsmath}
\usepackage{amssymb}
\usepackage{color}
\usepackage{bbding}
\usepackage[linesnumbered,ruled]{algorithm2e}
\usepackage{booktabs}
\usepackage[dvipsnames]{xcolor}
\usepackage[dvipsnames]{xcolor}
\hypersetup{
    colorlinks=true,       
    linkcolor=red,          
    citecolor=NavyBlue,        
}

\newcommand{\Dmat}{{\boldsymbol D}}

\newcommand{\Fmat}[0]{{{\boldsymbol F}}}
\newcommand{\Gmat}[0]{{{\boldsymbol G}}}

\newcommand{\Mmat}[0]{{{\boldsymbol M}}}

\newcommand{\Ymat}[0]{{{\boldsymbol Y}}}

\newcommand{\fv}{\boldsymbol{f}}

\newcommand{\gv}[0]{{\boldsymbol{g}}}

\newcommand{\vv}{\boldsymbol{v}}

\newcommand{\yv}{\boldsymbol{y}}

\newcommand{\Phimat}{\boldsymbol{\Phi}}

\begin{document}

\title{A Simple and Efficient Reconstruction Backbone for Snapshot Compressive Imaging}


\author{Jiamian Wang \and Yulun Zhang \and Xin Yuan \and Yun Fu \and Zhiqiang Tao}


\institute{
Jiamian Wang \at
Department of Computer Science and Engineering, Santa Clara University, USA \\
\email{JWang16@scu.edu}           
\and
Yulun Zhang \at
ETH Z\"{u}rich, Switzerland \\
\email{yulun100@gmail.com} 
\and
Xin Yuan \at
Westlake University, Hangzhou, Zhejiang, China \\
\email{xyuan@westlake.edu.cn}   
\and
Yun Fu \at
Department of ECE and Khoury College of Computer Science, Northeastern University, USA \\
\email{yunfu@ece.neu.edu}           
\and
Zhiqiang Tao \at
Department of Computer Science and Engineering, Santa Clara University, USA \\
\email{ztao@scu.edu}           
}

\date{Received: date / Accepted: date}

\maketitle

\begin{abstract}
The emerging technology of snapshot compressive imaging (SCI) enables capturing high dimensional (HD) data in an efficient way. It is generally implemented by two components: an optical encoder that compresses HD signals into a single 2D measurement and an algorithm decoder that retrieves the HD data upon the hardware-encoded measurement. Over a broad range of SCI applications, hyperspectral imaging (HSI) and video compressive sensing have received significant research attention in recent years. Among existing SCI reconstruction algorithms, deep learning-based methods stand out as their promising performance and efficient inference. However, the deep reconstruction network may suffer from overlarge model size and highly-specialized network design, which inevitably lead to costly training time, high memory usage, and limited flexibility, thus discouraging the deployments of SCI systems in practical scenarios. In this paper, we tackle the above challenges by proposing a simple yet highly efficient reconstruction method, namely stacked residual network (SRN), by revisiting the residual learning strategy with nested structures and spatial-invariant property. The proposed SRN empowers high-fidelity data retrieval with fewer computation operations and negligible model size compared with existing networks, and also serves as a versatile backbone applicable for both hyperspectral and video data. Based on the proposed backbone, we first develop the channel attention enhanced SRN (CAE-SRN) to explore the spectral inter-dependencies for fine-grained spatial estimation in HSI. We then employ SRN as a deep denoiser and incorporate it into a generalized alternating projection (GAP) framework -- resulting in GAP-SRN -- to handle the video compressive sensing task. Extensive experimental results on hyperspectral and video datasets demonstrate the state-of-the-art performance, high computational efficiency, and flexibility of the proposed SRN on two SCI applications. Code and pre-trained models are available at \url{https://github.com/Jiamian-Wang/HSI_baseline}.


\end{abstract}

\section{Introduction}\label{sce: intro}

\begin{figure*}[t]
\begin{center}
    \begin{subfigure}[b]{0.48\textwidth}
         \centering
         \includegraphics[width=1\linewidth]{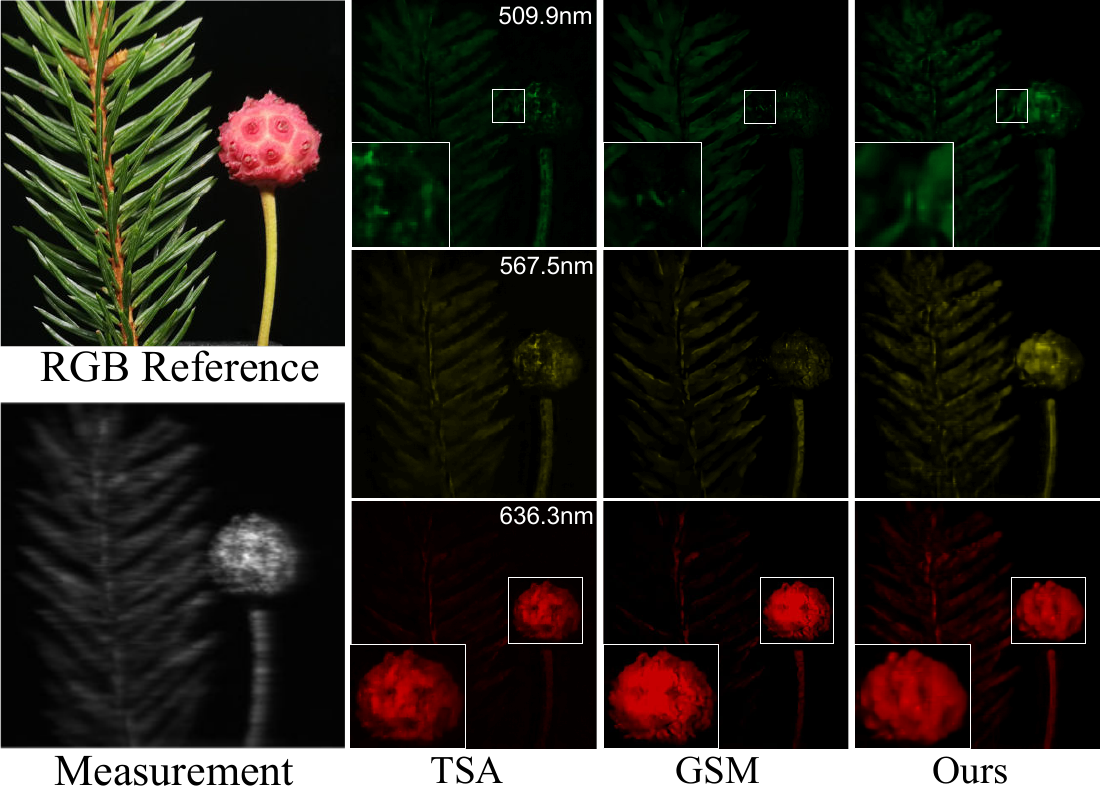}
         \caption{HSI reconstruction}
         \label{fig: spectral_real_cover}
    \end{subfigure}
    \hfill
    \begin{subfigure}[b]{0.48\textwidth}
         \centering
         \includegraphics[width=1\linewidth]{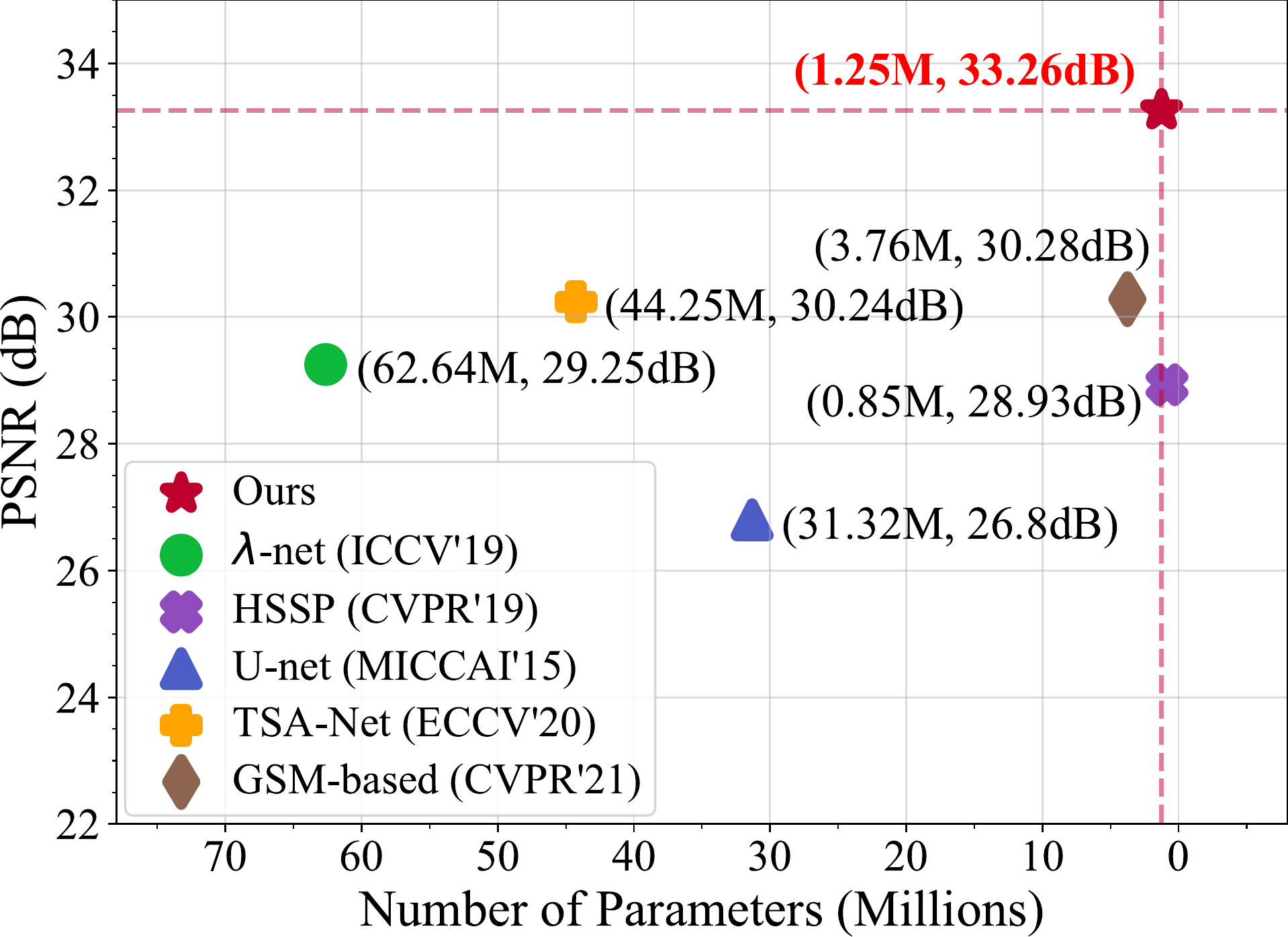}
         \caption{Model comparison}
         \label{Fig: scatter}
    \end{subfigure}
\end{center}
\vspace{-3mm}
\caption{(a) Real HSI reconstruction of TSA-Net, GSM and the proposed method. The RGB reference (top-left) is shown to demonstrate the color. By comparison, the proposed method perceptually outperforms the other two on exampled wavelengths. 
(b) Comparison between different methods in terms of PSNR ($\uparrow$) and the model size. ($\downarrow$).}
\vspace{-3mm}
\end{figure*}

Snapshot compressive imaging (SCI) ~\cite{yuan2021snapshot} captures the high-dimensional (HD) data in a hardware-encoding and software-decoding way. It is composed of an optical encoder to compress HD signals into 2D measurements and an algorithm decoder to retrieve the original signals with the hardware-encoded measurements. SCI systems are generally bandwidth/memory efficient and power-saving~\cite{yuan2021snapshot,llull2013coded,wagadarikar2008single} and thus widely used in different scenarios, \textit{e.g.}, temporal imaging~\cite{llull2013coded,yuan2016structured}, light field microscopy~\cite{he2020snapshot}, etc. Among various SCI applications, hyperspectral imaging and video compressive sensing have received significant research attention~\cite{Liu18TPAMI,Meng20ECCV_TSAnet,Miao19ICCV,meng2021self} from the computer vision community in recent years.

Hyperspectral imaging (HSI) obtains multi-channel images which store visual primitives at discrete spectral wavelengths~\cite{plaza2009recent}. By conveying rich information, hyperspectral images could better describe the nature of scenes than traditional RGB images and exhibit advantages in a wide range of scenarios, such as object detection~\cite{kim20123d,xu2015anomaly}, remote sensing~\cite{borengasser2007hyperspectral,melgani2004classification,yuan2017hyperspectral}, etc. In SCI, the hyperspectral data is encoded along the spectral axis and compressed into a measurement~\cite{Yuan2021_SPM}. This could be done by the coded aperture snapshot spectral imaging (CASSI) system~\cite{Meng20ECCV_TSAnet,Wagadarikar08CASSI}, which is a mainstream optical design among existing works~\cite{Wagadarikar08CASSI,Patrick13OE,Wagadarikar09CASSI,Yuan15JSTSP}.

Given the measurements by CASSI, many reconstruction algorithms~\cite{Meng20ECCV_TSAnet,Miao19ICCV,Wang_2020_CVPR,Wang17PAMI,Wang19TIP,Yuan16ICIP_GAP} have been proposed to retrieve hyperspectral images (Fig.~\ref{fig: spectral_real_cover}). For example, the optimization-based iterative methods~\cite{Liu18TPAMI,Bioucas-Dias2007TwIST} obtain the estimations in modulated signal domains constrained by pre-defined regularizers~\cite{wang2016adaptive,meng2020gap}.  They can be concatenated with distinct optical encoders without the re-training procedure and potentially own high robustness to modulation patterns and noise. However, they usually suffer from a relatively high time complexity and low-fidelity performance~\cite{meng2020gap}. The recent deep learning-based methods~\cite{Meng20ECCV_TSAnet,Miao19ICCV,Wang19TIP,huang2021deep} alleviate above issues by developing end-to-end reconstruction networks to directly approximate the solution of the inverse problem for SCI (\emph{i.e.}, optical-encoded measurements $\rightarrow$ hyperspectral images). While existing deep networks have achieved great success by yielding high-fidelity results with fast inference speed, their overlarge model size (considering limited hyperspectral images) and highly-specialized (only applicable to the HSI problem) network designs could lead to sub-optimal performance (see Fig.~\ref{Fig: scatter}) and narrow the application of a single reconstruction model on various SCI problems. To this end, the question \textit{whether there exists a network design with competitive performance, small model size, and simple structure} raises for the emerging field of SCI.

In this work, we propose a simple yet highly-efficient reconstruction model, termed as stacked residual network (SRN), to address the above challenge (Fig.~\ref{Fig: framework}(c)). While the residual learning~\cite{he2016deep} has been used in the previous HSI works~\cite{Meng20ECCV_TSAnet,Miao19ICCV,Wang_2019_CVPR}, we revisit it from two insights -- nested structure and spatial-invariance. Specifically, we implement a nested network structure using multiple local skip connections governed by a global one, which well balances the network depth and limited training data. On the other hand, we design the network with a spatial-invariant learning property to linearly enlarge the receptive field towards high-fidelity reconstructions. Empirically, the proposed SRN achieves state-of-the-art performance with negligible amount of parameters compared with previous methods. Moreover, the computational operations of SRN could be further reduced by utilizing the downsampling/upsampling rescaling pairs for the HSI reconstruction network setup. 

The proposed SRN is a versatile backbone that can be easily extended with advanced modules and is applicable to various SCI applications (\emph{e.g.}, HSI and video compressive sensing). Particularly, we propose a novel channel attention~\cite{hu2018squeeze,zhang2021accurate} enhanced SRN (CAE-SRN) to achieve a better computation/performance trade-off for the HSI problem. The channel attention provides more clues for spatial detail restoration from the scope of the spectral dimension by considering the channel-inter-dependencies, compensating for the missing details without overburdening the network. By the experiment, the proposed CAE-SRN clearly improves over SRN with the rescaling pairs included (see Table~\ref{Tab: model size}).

Besides HSI, we also apply the proposed backbone SRN in another mainstream SCI application -- video compressive sensing. Similarly, the sensing procedure also follows a encoding-decoding fashion. The coded aperture compressive temporal imaging (CACTI)~\cite{llull2013coded} is a representative SCI encoder to compress the video frames into a 2D measurement. For the decoding, existing methods can be divided into three categories. 1) The optimization-based iterative methods~\cite{Liu18TPAMI,yuan2014low,yuan2016generalized,ADMMnet2016NIPS} solve the inverse problem by designing novel regularizers and solvers. 
2) Deep reconstruction methods~\cite{cheng2020birnat,qiao2020deep} directly learn the mapping from a 2D measurement to temporal frames. For example, the recent proposed BIRNAT~\cite{cheng2020birnat} is capable to retrieve a 256$\times$256$\times$8 video data within 200ms and perceptually outperforms previous counterparts.
3) The hybrid methods that combine iterative algorithms and deep networks, \emph{i.e.}, deep unfolding~\cite{meng2020gap,yang2014video} and plug-and-play (PnP)~\cite{qiao2020deep,qiao2020snapshot} models.

However, bearing the goal of \textit{faithfully retrieving high-speed videos in real-time}, existing methods may show limitations in one way or another. On the one hand, the optimization-based iterative methods are a bit too slow, \emph{i.e.}, DeSCI requires over 6000s to reconstruct a 256$\times$256$\times$8 data. On the other hand, deep reconstruction networks are efficient but cannot comply with the general speed requirement, \emph{i.e.}, 30 fps, due to the large model size. In contrast, the hybrid method appears to be a promising direction. For example, the recent proposed GAP-net~\cite{meng2020gap} yields competitive performance with tens of times faster inference speed ($<$10ms) than BIRNAT~\cite{cheng2020birnat} on the same platform. While the GAP-net serves as one of the fastest SCI decoding algorithms (allowing $>$60 fps video retrieval theoretically), its deep reconstruction network usually sets a performance bottleneck due to the lack of lightweight backbones. In light of this, we develop GAP-SRN for the video compressive sensing task by incorporating SRN into the generative alternating projection~\cite{meng2020gap,Liao14GAP} framework. Specifically, SRN works as a ``deep denoiser'' for GAP, and its simple structure and efficient property facilitate the iterative optimizing procedure. By unfolding multiple SRN entities, the proposed GAP-SRN could explicitly exploit the temporal and spatial correlations in modulated signal domains.

To sum up, we provide a new deep reconstruction backbone for SCI in the spectral and video contexts. We summarize the contribution of this work in four-folds.
\begin{itemize}
\setlength\itemsep{0em}
    \item A simple yet highly-efficient SCI reconstruction network is provided by revisiting  the {residual learning} strategy, which could be used as a faithful {reconstruction backbone} for further modulations.
    \item Based on the proposed backbone, a novel CAE-SRN has been provided for HSI, which effectively retrieves the visual primitives by exploiting the {spectral inter-dependencies}. Besides, the model not only demonstrates high robustness to optical encoders, but also, to our best knowledge, empowers high-resolution spectral data retrieval for the first time.
    \item By integrating the proposed backbone into the generalized alternating projection (GAP) framework, we proposed a novel GAP-SRN that iteratively conducts the linear projection and SRN-based signal estimation for the task of the video compressive sensing. Owning to the fast inference and effectiveness of the proposed SRN, we make a solid step toward real-time high-speed video reconstruction. 
    \item Extensive experimental results on both hyperspectral and video datasets demonstrate the state-of-the-art performance of the proposed methods upon SRN and reveal the versatility of this proposed backbone on both popular applications of SCI.  
    
\end{itemize}

The organization of this paper is as follows. In Section~\ref{sec: related work}, the related works of SCI on both applications are outlined. In Section~\ref{sec: method}, we introduce the mathematical background of SCI and propose the backbone, stacked residual network (SRN). We further describe the CAE-SRN for HSI and GAP-SRN for video reconstruction. In Section \ref{sec: experiments}, we perform extensive experiments to verify the effectiveness of proposed methods on above applications, respectively. Finally, we draw the conclusion and discuss the future work in Section~\ref{sec: conclusion}.

\section{Related Work \label{sec: related work}}
This study focuses on two popular applications hyperspectral imaging (HSI) and video compressive sensing based on spectral and video SCI systems, respectively.

\subsection{Spectral SCI}\label{subsec: spectral sci}
For spectral SCI, the coded aperture snapshot spectral imager (CASSI) is provided as a promising detecting instrument due to its convenience in optical alignment and simplicity in spectral projection~\cite{wagadarikar2008single}. It firstly encodes the hyperspectral data upon specific wavelengths via multiple masks (\emph{i.e.}, shifted version of a single coded aperture). Then it projects the signal onto a 2D measurement via a disperser. Following this, many software-based reconstruction algorithms are proposed upon advanced compressive sensing (CS) algorithms~\cite{candes2006robust,donoho2006compressed}. 

Previously, iterative-based algorithms demonstrate promising reconstruction performance. To obtain desired results, diverse priors are employed as regularization terms, constraining the optimization space for the iteration. For example, ~\cite{reddy2011p2c2} solves the problem through wavelet-based regularization, the GPSR~\cite{Figueiredo07GPSR} applies the sparse prior for reconstruction. The TV priors are included in TwIST~\cite{Bioucas-Dias2007TwIST} and GAP-TV~\cite{Yuan16ICIP_GAP}. Among iterative-based algorithms, DeSCI~\cite{Liu18TPAMI} sets the state-of-the-art performance by proposing a novel framework under the flow of rank minimization. Despite the prosperity of iterative-based algorithms, challenges like long reconstruction time (\emph{i.e.}, DeSCI requires more than an hour for an 256$\times$256$\times$8 data reconstruction) and potentially unstable convergence still exist.

By comparison, recently well-explored deep learning (DL)-based methods are characterized by fast reconstruction (\emph{i.e.}, the inference time is always at second-level) and simultaneously, high retrieval performance. They seek to solve the problem from different perspective of views. The $\lambda$-net~\cite{Miao19ICCV} trains a dual-stage reconstruction model under the framework of the generative adversarial network. The TSA-Net~\cite{Meng20ECCV_TSAnet} explores the spectral and spatial self-attentions in an order-independent manner for the first time, achieving impressive reconstruction performance since proposed. The GSM~\cite{huang2021deep} method proposes a Maximum a Posterior (MAP) estimator upon learnable Gaussian Scale Mixture (GSM) prior, with included sub-modules embodied by deep neural networks. It yields the best performance among concurrent reconstruction algorithms. The prevalence of the DL techniques facilitate the wide deployment of SCI systems to diverse real-world applications. However, the heavy computation burden and long training time induced by the large model size prohibit such a development. Also, it is inconvenient to make further modifications on existing highly-specialized network structures, setting up barriers to the future works. 

To mitigate these problems, one practical solution is to integrate the advantages of deep neural networks and iterative-based methods. This leads to twofold popular frameworks: 1) deep unfolding (unrolling) methods, which conducts optimization iterations upon a series of concatenated deep neural networks, \emph{i.e.}, GAP-net~\cite{meng2020gap} and HSSP~\cite{Wang_2019_CVPR}. Since linear projections are conducted between the networks (stages), one can flexibly expand or shrink the whole framework. 2) Pluy-and-play (PnP) frameworks~\cite{meng2021self,Zheng20_PRJ_PnP-CASSI}, which treat the pre-trained neural networks as deep denoisers in the iterative-based algorithms. Both deep unfolding and PnP methods could be flexibly modulated.  They also simultaneously benefit from iterative operations and learning capacity of undetermined neural network settings. Related methods are expected to be relatively light weight (as compared in Fig.~\ref{Fig: scatter}, HSSP only contains less than 1M parameters) and demonstrate the robustness to masks~\cite{meng2020gap}. 

On the other hand, a more direct solution of improving the underlying flexibility and efficiency is to introduce lightweight and simple-structured networks as reconstruction backbones, which could be either solely used or conjointly integrated into other frameworks. A popular option is the U-Net~\cite{ronneberger2015u}. For example, the TSA-Net~\cite{Meng20ECCV_TSAnet} builds upon U-Net structure. The GSM ~\cite{huang2021deep} employing U-Net for the regularization parameter approximation and the prior local mean estimation. The GAP-net~\cite{meng2020gap} with U-Net leads to a pleasing results. However, the low-fidelity performance (\emph{i.e.}, $<$27dB by PSNR) of solely-used U-Net reveals that its inherently limited ability in processing hyperspectral data.

Notably, the residual learning~\cite{he2016deep} strategy beneath the U-Net structure serves as another popular schema for deep network design~\cite{Miao19ICCV,meng2020gap,Wang_2019_CVPR} in SCI. However, none of existing works truly exploit the underlying potential of the schema. Based on this observation, we take advantage of the residual learning by further proposing a nested structure with a spatial-invariant characteristic.

\subsection{Video SCI}\label{subsec: video sci}
For video SCI, the coded aperture compressive temporal imaging (CACTI) compresses the high-speed scenes at a lower capture rate~\cite{llull2013coded}. Firstly, CACTI uses an objective lens to take the temporal frames. They are further encoded by physical masks with different spatial patterns displayed on the digital micromirror device (DMD). Finally, the monochrome/color CCD detects the 2D measurement that contains the information of the corresponding compressed signal. 

For the video frame reconstruction, existing algorithms can be similarly categorized into three groups: 1) model-based methods, among which the DeSCI~\cite{liu2018rank} benefits from the weighted nuclear norm minimization (WNNM)~\cite{gu2014weighted} and alternating direction method of multipliers (ADMM) solver~\cite{boyd2011distributed}, delivering the most promising performance. 2) Deep neural networks directly learn a straightforward mapping from the measurement to video frames. For example, the E2E-CNN~\cite{qiao2020deep} supports millisecond-level retrieval speed. The BIRNAT~\cite{cheng2020birnat} employs a recurrent neural network to explore the correlation between the video frames. Also, empowered by adversarial training, this novel method perceptually outperforms existing methods. 3) Unfolding and PnP-based methods form another popular stream. Recently, a PnP-FFDNet~\cite{yuan2020plug} that takes the FFDNet~\cite{zhang2018ffdnet} as the denoiser achieves promising performance with second-level retrieval time on 256$\times$256$\times$8 video data reterieval. However, the capacity of the pre-trained denoiser sets the bottleneck for this method. By comparison, the GAP-net~\cite{meng2020gap} demonstrates more descent results by flexibly integrating proper denoiser among several candidates including AutoEncoders, U-nets and ResNets etc., by training from scratch. 

\section{Methodology \label{sec: method}}

\begin{figure*}[tp] 
\centering 
\includegraphics[width=1\textwidth]{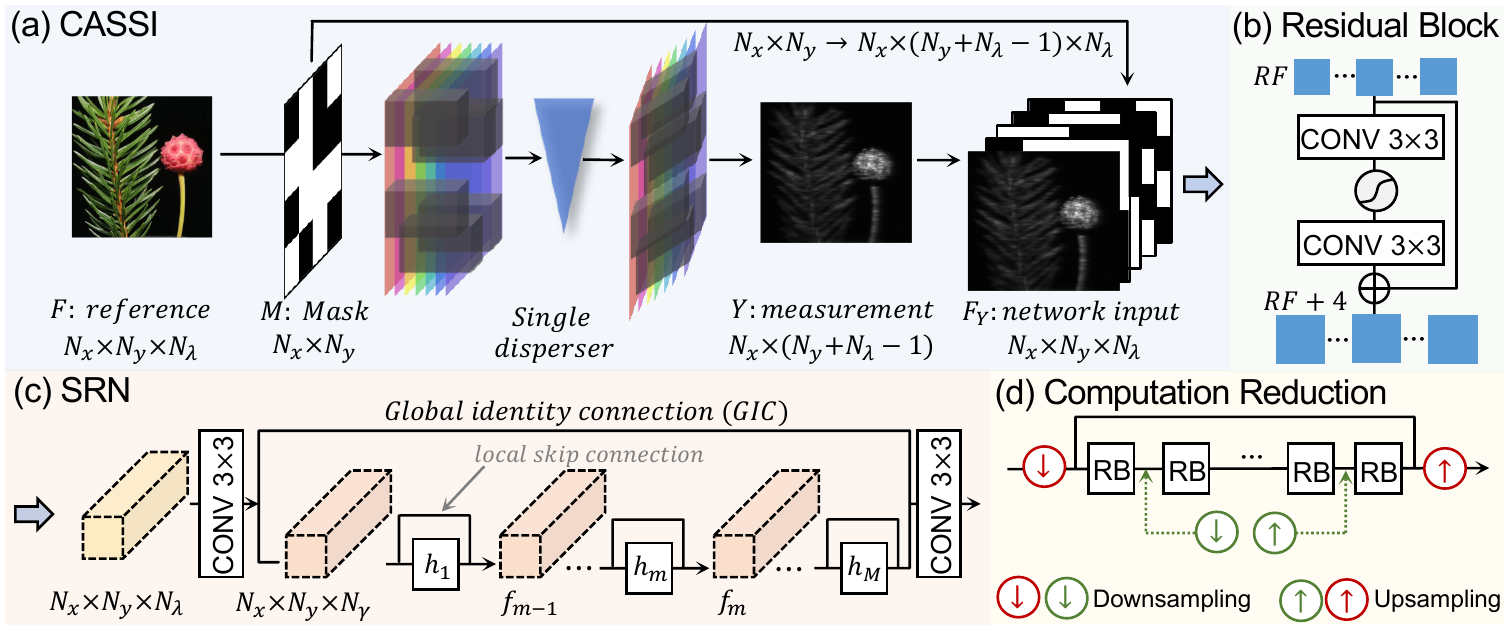}\vspace{-2mm}
\caption{CASSI compression and software-based reconstruction for spectral SCI. Note that we exchange $N_{c}$ with  $N_{\lambda}$. (a) In CASSI, a 2D coded aperture (mask) $\Mmat$ encodes the HSI signal $\Fmat$ at specific wavelengths. A disperser shears the signal along the y-axis. The Network input $\Fmat_Y$ is initialized by measurement and shifted version of the mask. (b) Structure of the residual block. (c) Architecture of the SRN. It is characterized by nested residual learning and spatial-invariant learning. (d) Computation reduction through rescaling pairs, which can be flexibly plugged into the network. we use 2-paired rescaling as an example.} 
\label{Fig: framework}
\vspace{-5mm}
\end{figure*}

\subsection{Mathematical Model of SCI \label{subsec: mathmatical}}
\textbf{Forward Model}.
The forward encoding process of SCI can be expressed via a unified mathematical model. In this work we take the 3D data (spectral data or video frames) as an example. Given the cube $\Fmat \in {\mathbb R}^{N_x \times N_y \times N_{c}}$ where $N_x$, $N_y$, and $N_{c}$ represent the height, width and number of channels (frames), respectively. The 2D measurement $\Ymat$ is obtained by modulating data cube with masks $\Mmat \in {\mathbb R}^{N_x \times N_y \times N_{c}}$ along the target dimension as
\begin{equation}
\Ymat = \sum\nolimits_{n_{c}=1}^{N_{c}} \Fmat (:,:, n_{c}) \odot \Mmat (:,:, n_{c}) + \Gmat, \label{eq: unified model}
\end{equation}
where $\Gmat$ denotes a 2D measurement noise and $\odot$ is the Hadamard product. Notably, the $N_c$ in Eq.~\eqref{eq: unified model} could refer to the number of wavelength ($N_{\lambda}$) in spectral SCI or the number of temporal frames ($N_{t}$) for compression in video compressive sensing. The forward model of SCI can be alternatively expressed in vector domain by 
$\fv$=${\rm vec}( [\fv^{(1)}\dots{\fv}^{(N_{c})}])$
where ${\fv}^{(n_{c})} = {\rm vec}( \Fmat (:,:, n_{c}))$ and similarly determining the sensing matrix 
\begin{equation}\label{eq: sensing matrix}
\begin{aligned}
\Phimat = \left[\Dmat_1, \dots, \Dmat_{N_{c}}\right], 
\end{aligned}
\end{equation}
where $\Dmat_{n_{c}} = {\rm Diag} ({\rm vec}(\Mmat (:,:, n_{c})))$. To this end, a vectorized forward model could be yield as
\begin{equation}\label{eq: vec model}
\yv = \Phimat {\fv} + \gv. 
\end{equation}

In the following, we will deploy this generalized forward model in CASSI and CACTI for hyperspectral imaging and video compressive sensing, respectively.

\noindent\textbf{CASSI for Spectral SCI}.
The spectral SCI is specifically associated with the the spectral data corresponding to numerous wavelengths. Regarding this, a prevailing spectral imager, CASSI, employs a fixed 2D coded aperture (mask) plus a disperser, which is equivalent to different masks, leading to a passive modulation. Here we use $N_{\lambda}$ exchange $N_{c}$. For aforementioned spectral data $\Fmat$, we firstly compute signal modulation implemented by the mask in a channel-wisely way as
\begin{equation}\label{eq: mask encoding}
{\Fmat}' (:,:,n_{\lambda}) = {\Fmat} (:,:,n_{\lambda}) \odot \Mmat^*,
\end{equation} 
where ${\Fmat}'\in {\mathbb R}^{N_x \times N_y \times N_{\lambda}}$ denotes the modulated signals, $\Mmat^* \in  {\mathbb R}^{N_x \times N_y}$ refers to a pre-defined physical mask, $n_{\lambda} \in [1,\dots, N_{\lambda}]$ indexes wavelengths.
By passing to a single disperser (SD), the $\Fmat'$ is tilted and sheared along the $y$-axis. Let $\Fmat'' \in {\mathbb R}^{N_x \times (Ny + N_{\lambda}-1) \times N_{\lambda}}$ be the tilted cube, and $\lambda_c$ be a reference wavelength, \emph{i.e.}, $\Fmat'[:,:,n_{\lambda_c}]$ works like an anchor image without shearing
and
\begin{equation}\label{eq: disperser modulation}
\Fmat'' (u,v, n_{\lambda}) = \Fmat'(x, y + d(\lambda_n - \lambda_c), n_{\lambda}),
\end{equation}
where $(u,v)$ locates the coordinate system on the detector plane, $\lambda_n$ denotes the $n_\lambda$-th channel, $\lambda_c$ refers to the anchored wavelength, and $d(\lambda_n -\lambda_c)$ represents a spatial shift of the $n_\lambda$-th channel. 
We further have
\begin{equation}\label{eq: measurement obtain}
\Ymat = \sum\nolimits_{n_{\lambda}=1}^{N_{\lambda}}  \Fmat'' (:,:, n_{\lambda}) + \Gmat,
\end{equation}
where $\Ymat \in {\mathbb R}^{N_x \times (Ny + N_{\lambda}-1)}$ is the measurement and $\Gmat\in {\mathbb R}^{N_x \times (Ny + N_{\lambda}-1)}$ denotes the measurement noise. Combining the mask encoding in Eq.~\eqref{eq: mask encoding} and shifting in Eq.~\eqref{eq: disperser modulation}, we express the passive modulation as 
\begin{equation}\label{eq:M_F}
\begin{aligned}
&\Mmat (u,v, n_{\lambda}) = \Mmat^*(x, y + d(\lambda_n - \lambda_c)), \\
&\tilde{\Fmat} (u,v, n_{\lambda}) = \Fmat(x, y + d(\lambda_n - \lambda_c), n_{\lambda}).
\end{aligned}
\end{equation}
By substituting original data cube in Eq.~\eqref{eq: unified model} with $\tilde{\Fmat}$, the measurement $\Ymat$ obtained in Eq.~\eqref{eq: measurement obtain} becomes
\begin{equation}\label{eq:sensing_Matrix}
\Ymat = \sum\nolimits_{n_{\lambda}=1}^{N_{\lambda}}   \tilde{\Fmat} (:,:, n_{\lambda})  \odot \Mmat (:,:, n_{\lambda}) + \Gmat. 
\end{equation}

The subsequent vectorization of $\Ymat$ exactly follows the unified model. Notably, the sensing matrix $\Phimat$ given in Eq.~\eqref{eq: sensing matrix} has a very special structure that most existing well-developed CS theories~\cite{Donoho06ITT,Candes06ITT} cannot easily handle, \emph{i.e.}, $\Phimat$ is not only a fat matrix but also highly sparse –– it is composed of $N_{\lambda}$ concatenated diagonal matrices with at most $nN_{\lambda}$ nonzero elements.

Given the measurement $\yv$ captured by the SD CASSI system, the reconstruction is to solve $\fv$ used in Eq.~\eqref{eq: vec model}, falling in the vein of inverse problem~\cite{Yuan2021_SPM}. Since the sparse sensing matrix $\Phimat$ is determined by the masks, deep reconstruction networks like TSA-Net~\cite{Meng20ECCV_TSAnet} firstly discretizes the input measurement by shifted version of masks. Following the same procedure, we assume that the underlying mapping is more inclined to identity relationship, yielding a residual learning skeleton.

\noindent\textbf{CACTI for Video SCI}.
In contrast with the spectral SCI, video SCI refers to capturing high-speed video frames at a lower capture rate by the compressive imaging system, \emph{i.e.}, CACTI. Different from the CASSI, CACTI directly applies shifting masks or variant patterns on the digital micromirror device (DMD), resulting in an activate modulation. Due to this, it acts as a low-power imager by skipping code transmission during the signal modulation~\cite{llull2013coded}. As the CACTI and CASSI functionally work in the same way, one can directly refer to the generalized forward model for video SCI.

\subsection{A Simple SCI Reconstruction Backbone by SRN} \label{subsec: SRN}

In this section, we introduce the proposed stacked residual network (SRN) in detail. Despite the simple structure, it effectively exploits the potential of nested residual learning (NRL) and benefits from the characteristic of spatial-invariant learning. By employing rescaling pairs, we make a further step on computation reduction, yielding a light-weight and highly-efficient model.

For better illustration, we introduce the backbone under the application of spectral SCI. Firstly the hyperspectral signal is captured by the optical apparatus with a SD CASSI system~\cite{Meng20ECCV_TSAnet}. As demonstrated in Fig.~\ref{Fig: framework} (a), the 3D signal corresponding to a given real-world (or synthetic) scene is compressed by a 2D measurement $\Ymat$. 
We use ${\Fmat}_{Y} \in {\mathbb R}^{N_x \times Ny \times N_{\lambda}}$ as the initialized input for the network and define its $n_{\lambda}$-th channel as  
\begin{equation}\label{eq: input initialization}
    {\Fmat}_{Y}[:,:,n_{\lambda}] :=shift(\Mmat_{n_{\lambda}} \odot \Ymat),
\end{equation}
where $shift$ modulates the product result back to original spatial dimension of HSI data.
Overall, the reconstruction mapping is obtained by
\begin{equation}
    f_{SRN}(\cdot):\Fmat_{Y} \rightarrow \widehat{\Fmat},
\end{equation}
where $\widehat{\Fmat}$ denotes the reconstruction result. Essentially, the proposed SRN $f_{SRN}(\cdot)$ learns to fit $\Phimat^{-1}$ expressed in a vector space given in Eq.~\eqref{eq: vec model}.

As shown in Fig.~\ref{Fig: framework} (c), the majority of the SRN is the nested residual learning module. This requires the residual output shares the same dimension as the network input in the spectral domain. Therefore, directly employing the NRL poses a large limitation to the network width and potentially compromise the learning capacity~\cite{zagoruyko2016wide}. Therefore, we create feature embedding with manipulated spectral channels by implementing \texttt{CONV} layers: ${\mathbb R}^{N_x \times Ny \times N_{\lambda}} \rightarrow  {\mathbb R}^{N_x \times Ny \times N_{\gamma}}$. Conversely, spectral channels are recovered after the NRL.

\subsubsection{Nested Residual Learning}
It is widely known that the network depth contributes to the modeling ability~\cite{szegedy2017inception}, but poses obstacles to the training on the other hand~\cite{zagoruyko2016wide}. Putting emphasize on nested residual learning upon global identity connection and local skip connections (LSCs) ensures the flexible gradient propagation within the network~\cite{he2016identity}.

For the residual mapping part, we concatenate multiple Residual Blocks (RBs) governed by local skip connections separately. By doing this, one can customize the network structure adapting to different SCI applications, which makes the SRN differentiate from other deep models with deterministic architectures. No matter how many RBs are employed, the network remains well organized and transductive. The feed-forward operation between RBs is proceeded as
\begin{equation}\label{eq: rb-feedforward}
    f_m = H_m(f_{m-1})=h_m(f_{m-1})+f_{m-1},
\end{equation}
where $H_m(\cdot)$ denotes the residual learning given by a CONV-ReLU-CONV structure with a local skip connection, \emph{i.e.}, $H_m(x)$ = $\mathrm{CONV}(\mathrm{ReLU}(\mathrm{CONV}(x))) + x$, and $f_{m-1}$ represents the output of the $(m-1)$th Residual Block. Supported by experimental results, we remove the Batch Normalization~\cite{ioffe2015batch} in RBs. The whole stacked residual part could be represented as
\begin{equation}\label{eq: fm-rbs}
\begin{aligned}
    f_{M} = H_{M}(H_{M-1}( \dots H_2(H_1(f_0)) \dots)),
\end{aligned}
\end{equation}
where $f_0$ is given as the feature embedding before NRL. Thus, we could treat all the RBs in Eq.~\eqref{eq: fm-rbs} as a spatial / spectral-invariant mapping $f_{M}: \mathbb{R}^{N_x \times N_y \times N_\lambda}$ $\rightarrow$ $\mathbb{R}^{N_x \times N_y \times N_\lambda}$, and define the reconstruction model as 
\begin{equation}\label{eq: f-res}
\begin{aligned}
f_{SRN} \!=\! \textup{CONV}(f_M(\textup{CONV}(\Fmat_{Y}))+\textup{CONV}(\Fmat_{Y})).
\end{aligned}
\end{equation}
Overall, such an simple and well-organized and network structure shares similar principles with existing residual-based reconstruction models but is characterized by intertwined residual hierarchies, thus fully exploits the potential of the residual learning schema.

\subsubsection{Spatail-invariant Learning}
\begin{figure}[t]
\begin{center}
\includegraphics[width=1\linewidth]{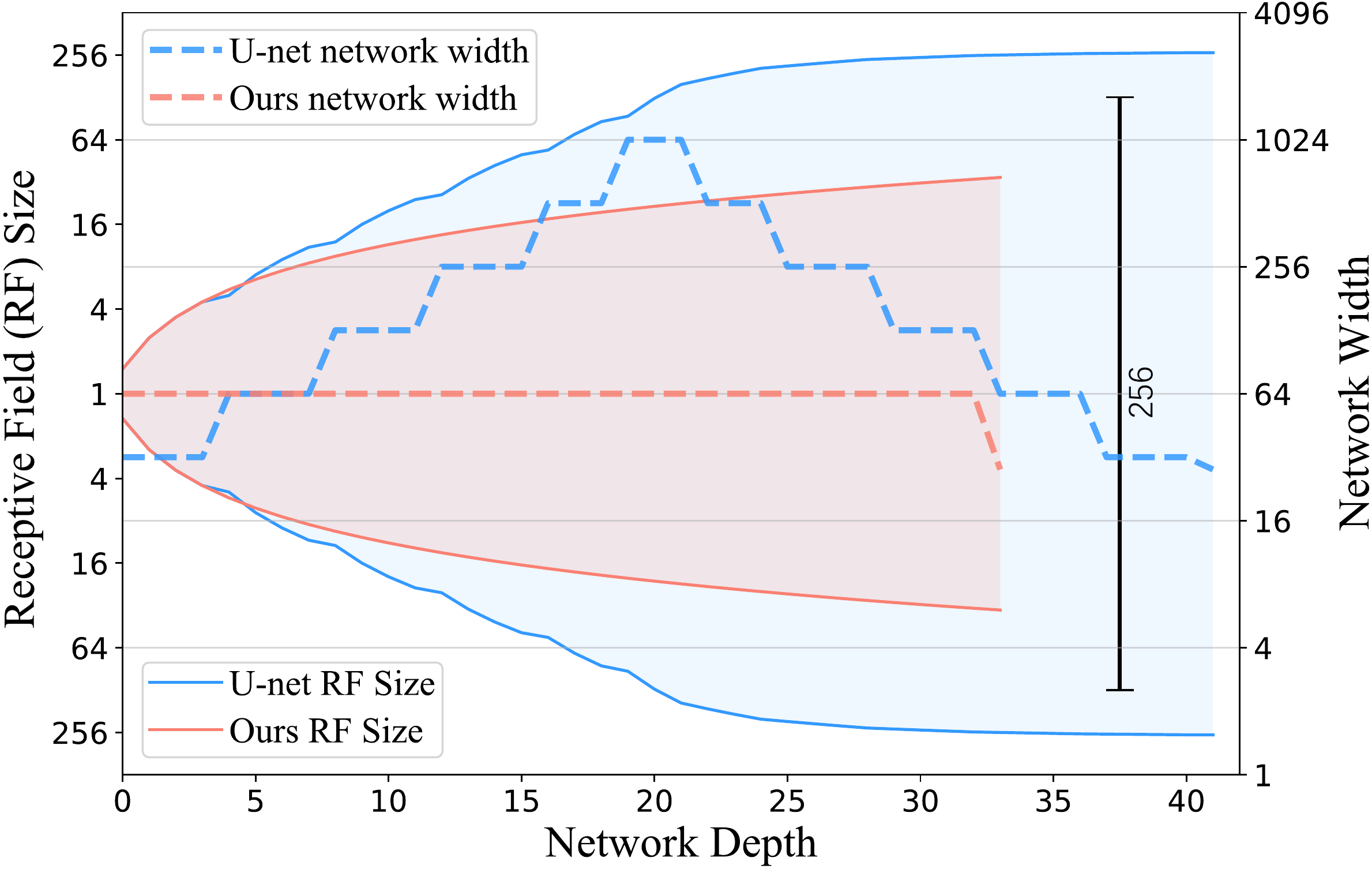}\vspace{-3mm}
\end{center}
\caption{Receptive field (RF) size and network width comparison plotted in Log-scale (U-Net~\cite{Miao19ICCV} v.s. Ours). The RF of our method could be linearly extended with a salable structure. Compared with that of the U-Net (\emph{i.e.}, 532$\times$532), we find RF of 70$\times$70 is sufficiently large given input of 256$\times$256 (black bar). Besides, the network width of SRN simply remains invariant.}
\label{fig: unet_vs_ours}
\vspace{-3mm}
\end{figure}

The usage of multiple Residual Blocks not only enables arbitrary network depth, also benefits the architecture customization from the perspective of receptive field (RF), which can be generally enlarged by stacking more layers. The RF should be sufficiently large to capture the semantic information for tasks like image classification and segmentation~\cite{luo2016understanding}. However, trade-offs need to be made considering the input image size, potential size of relevant area and computation burden. One can easily measure the RF of the SRN since the area is linearly increased with more RBs. This is implemented by the same kernel size and padding operation in Residual Blocks (Fig.~\ref{Fig: framework} (b)), leading to the \textit{spatial-invariant learning} -- the spatial output size of all the RBs are the same in the proposed SRN backbone.
Fig.~\ref{fig: unet_vs_ours} shows the comparison of RF between SRN and the U-net backbone. Given the 256$\times$256$\times$28 HSI data, the consequent RF of SRN is 70$\times$70, which is proven to be sufficiently large for a compelling performance. While that of the U-Net is over 500$\times$500, which turns out that barely contributes to the reconstruction performance but takes more computational resource (see Table~\ref{Tab: model size}).

Notably, the network width (channels of network embedding) $N_{\gamma}$ also remains invariant across the NRL module, as demonstrated in Fig.~\ref{fig: unet_vs_ours}. 
Particularly, we set the $N_{\gamma}$ to be much larger than the number of initial spectral channels $N_{\lambda}$ (\emph{i.e.}, 64 v.s. 28) to increase the spectral-wise redundancy of intermediate embedding.

\subsubsection{Computational Efficiency Enhancement}
In this work, we measure the computation burden by floating point operations (FLOPs), \emph{i.e.}, the amount of additions and multiplications. Larger model with larger input spatial size generally end up with high FLOPs. In addition, both the larger model size and the higher FLOPs indicate more GPU memory usage. Despite the small model of proposed SRN, the computation included is still significant. As compared in Table~\ref{Tab: model size}, the SRN taking 3\% parameters of TSA-Net contains similar computation instead. This indicates the underlying computation time and resource occupation still remain considerable, which limits the future deployment of SRN in diverse scenarios. Take the high-resolution HSI (HR-HSI) reconstruction as an example. Existing deep reconstruction networks set demanding requirements to the GPU memory, therefore rarely visualizing reconstruction from HR-HSI over 1000 pixel scale. Aiming at efficient and power-saving reconstruction, we seek to reduce computation burden of SRN.

We consider simply introducing rescaling pairs for the objective. It essentially contains a downsampling operation for a leap upon receptive fields and an upsamping operation exactly doing the inverse. By processing feature maps with much smaller spatial size inclusively, one can diminish the computation without disturbing key learning elements (\emph{i.e.}, model size, network depth). Another advantage is that the pairs can be flexibly plugged in the salable network structure, providing more options over network design. We implement the downsampling operation with \texttt{CONV} layer at a specific stride. For upsampling, we use famous \textit{PixelShuffle}~\cite{shi2016real} instead of traditional interpolations or deconvolution. Specifically, the operation can be expressed as $\mathrm{R}^{H \times W \times Cr^2} \rightarrow \mathrm{R}^{Hr \times Wr \times C}$ with upsampling scale $r$.

As shown in Fig.~\ref{Fig: framework} (d), rescaling pairs can be directly inserted between Residual Blocks. The larger span between down/up-sampling, the more computation mitigated. In this work, we try two different cases. Based on the original SRN (dubbed as \texttt{SRN(v1)}), we let the rescaling pair with scale 2 right outside the NRL module. Based on this variant, we further introduce one more rescaling pair in the NRL, \emph{i.e.}, includes $K$ RBs in the middle, yielding \texttt{SRN(v3)}. From experiment results, we observe that three variants lead to the state-of-the-art/competitive performance under simulation data (refer to Table~\ref{Tab: psnr}), while the performance will descend with more rescaling pairs. This is due to the lossy compression by the downsampling. In the following we will allivate this problem via channel attention technique.

\subsection{CAE-SRN for Spectral SCI}\label{subsec: CAE+SRN}
How to make up for the informative detail loss by rescaling pairs? Previous works provide a possible solution by revealing main channel features~\cite{hu2018squeeze}. That is to say, better reconstruction can be obtained by differentiating the importance of channels, under missed visual descriptions. In single image denoising, empirical evidence indicates that the attention guided scaling~\cite{zhang2021accurate} alleviates the obstacle of informative loss. Inspired by this, we signifie the channel importance by channel attention enhancement (CAE) module into Residual Blocks.

\begin{figure}[t]
\begin{center}
\includegraphics[width=0.8\linewidth]{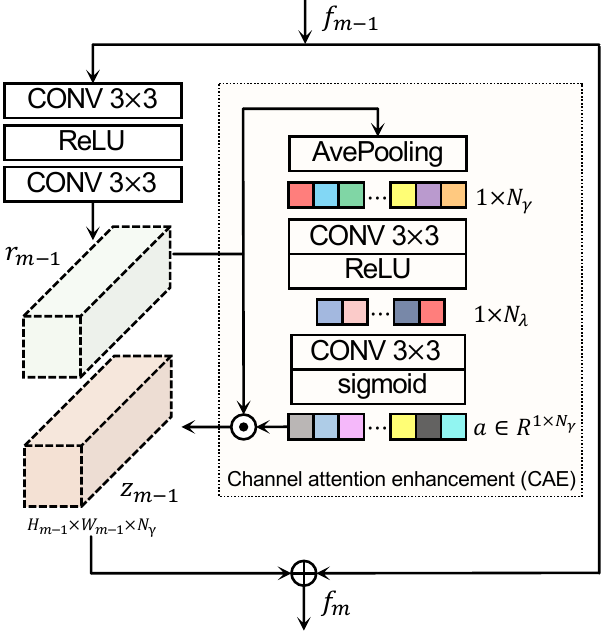}\vspace{-5mm}
\end{center}
\caption{Details of channel attention enhancement (CAE). The Residual Block governs the CAE module. The CAE vector $a$ is computed by exploiting channel informative details (via global average pooling) and channel dependencies learned by nonlinear representation (via \texttt{CONV}s and activation functions). }
\label{fig: eca}
\vspace{-6mm}
\end{figure}

The main idea is to explicitly discriminate $N_{\gamma}$ channels according to the inter-dependency and informative abundance~\cite{zhang2018image}. As shown in Fig.~\ref{fig: eca}. Given the input $f_{m-1} \in \mathrm{R}^{N_x \times Ny \times N_{\gamma} }$ of the $m$-th Residual Block, $r_{m-1}$ represents the original residual output as $r_{m-1}=\texttt{CONV}(\texttt{ReLU}(\texttt{CONV}(f_{m-1}))$. We initialize the channel attention $a$ by global average pooling as
\begin{equation}\label{eq: attention vec}
a[n_{\gamma}]=\frac{1}{H_{m-1} W_{m-1}}r_{m-1}[:, :, n_{\gamma}],
\end{equation}
where $n_{\gamma}=1,...,N_{\gamma}$. Next, we fuse channel informative degrees and determine CAE according to the dependency, implemented by
\begin{equation}
    a \leftarrow \texttt{CONV-sigmoid}(\texttt{CONV-ReLU}(a)).
\end{equation} 
The attention makes effect by multiplication
\begin{equation}\label{eq: ECA multiply}
    z_{m-1}[:,:,n_{\gamma}]=a[n_{\gamma}] \odot r_{m-1}[:,:,n_{\gamma}],
\end{equation}
where $z_{m-1}[:,:,n_{\gamma}]$ is the CAE-residual. Finally, we obtain the output of RB by $f_m=f_{m-1} + z_{m-1}$. In original SRN, all residual channels are equally accounted in each RB. By channel attention enhancement, more possibilities could be exploited in a spectral-wise, potentially compensating for the spatial informative detail deficiency. Consistently, we indeed observe obvious performance boosts on \texttt{SRN(v2)} and \texttt{SRN(v3)} with CAE.

\noindent\textbf{Model Training.} The proposed CAE-SRN is trained to minimize a mean squared error (MSE) between the ground truth HSIs and reconstruction results
\begin{equation}
    \mathcal{L}_{MSE} (\phi)= \frac{1}{N} \sum_{n=1}^N ||\widehat{\Fmat}_n - \Fmat_n||^2,
\end{equation}
where $\widehat{\Fmat}$=$f_{\textup{CAE-SRN}}(\Fmat_{Y})$ is given in Eq.~\eqref{eq: f-res} and $N$ is the number of samples. $\phi$ denotes all the parameters.

For simulation, synthetic hyperspectral images are fed into the SD CASSI for input initialization according to Eq.~\eqref{eq: input initialization}. Therefore, the original 3D cube naturally becomes the ground truth (GT). 
For real-world HSI reconstruction, training is conducted upon the simulation training set plus imposing Gaussian noise on the measurements to mimic real-captured measurements. We uniformly sample the standard deviation (std) from $[0, 0.05]$, and employ the measurements generated by a real SD CASSI system to the trained model.

\subsection{GAP-SRN for Video SCI}\label{subsec: GAP-SRN}
As a versatile network design for SCI reconstruction, not only the proposed SRN enables outstanding retrieval performance in the application of hyperspectral imaging, it also could be effectively leveraged for video compressive sensing as a promising backbone. Specifically, we integrate the SRN into a popular unfolding framework of SCI, namely generalized alternating projection (GAP)~\cite{Liao14GAP}. 
Following the previous attempts of deep networks in GAP~\cite{meng2020gap}, the SRN plays role of a ``deep denoiser", modulating the signal to desired domains~\cite{yuan2021snapshot}. Such a combination brings about following advantages: 1) GAP framework enables competitive performance on video SCI~\cite{meng2020gap}. 2) Under the GAP framework, distinct network structures, \emph{i.e.}, AutoEncoder, U-Net, ResNet and DnCNN~\cite{zhang2017beyond} are previously utilized, providing adequate references for underlying comparison. 3) The performance of this compound method largely depends on the backbone being applied. While existing structures enables unsatisfactory results.  
4) Besides, the well-defined architecture of SRN brings advantage in signal modulation under GAP framework,  considering the denoising nature of the modulation and effectiveness of residual learning in denoising~\cite{zhang2017beyond}. 5) When jointly considered, the combination of SRN and GAP framework could explicitly abstract spatial \ spectral correlations in diverse signal domains.

\begin{figure}[t]
\begin{center}
\includegraphics[width=0.99\linewidth]{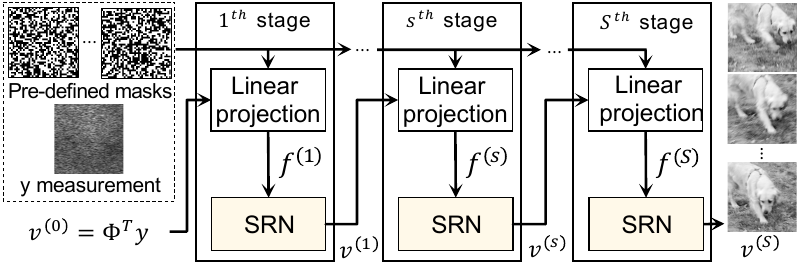}\vspace{-3.5mm}
\end{center}
\caption{Structure of GAP-SRN. It takes the measurement $\yv$ and pre-defined masks as input. The framework consists of $S$ stages. Each stage computes $\fv^{(s)}$ by Eq.~\eqref{eq: linear projection} and estimates $\vv^{(s)}$ by SRN. $\Phimat$ is determined by masks in Eq.~\eqref{eq: sensing matrix}. We train the model end-to-end. }
\label{fig: gap_srn}
\vspace{-3mm}
\end{figure}

Recall that the forward model given in Section~\ref{subsec: mathmatical} is applicable to the video SCI. Accordingly,  considering the mathematical model in Eq.~\eqref{eq: vec model}, GAP-net iteratively optimizes the signal, to solve the following objective 
\begin{equation}\label{eq: recon obj}
    \widehat{\fv}=\mathop{\arg\min}\limits_{\fv}\frac{1}{2}\| \yv-\Phimat {\fv}\|^2_2 + \pi \Omega(\fv),
\end{equation}
where $\Omega(\fv)$ denotes the regularization term constraining the signal domain. By investigating the augmented Lagrangian formulation~\cite{sun2016deep} of the objective, the GAP-net iteratively solves the Eq.~\eqref{eq: recon obj} in two steps: 1) linear projection and 2) solving nonlinear shrinkage function by a denoising network. As shown in Fig.~\ref{fig: gap_srn}, it contains $S$ repeated backbones as denoising networks concatenated by linear projections:
\begin{equation}\label{eq: linear projection}
    \fv^{(s+1)}=\vv^{(s)} + \Phimat^{T}(\Phimat \Phimat^{T})^{-1}(\yv-\Phimat \vv^{(s)}),
\end{equation}
where $\vv^{(s)}$ is an auxiliary variable  to express the network output of the $s$-th network, which prepares the input $\fv^{(s+1)}$ for the ($s$+1)-th network. Let $f_{\textup{SRN}}$ again represents the denoising mapping, the backbones essentially work as $\fv^{(s+1)}=f_{\textup{SRN}}(\vv^{(s)})$. The above two steps consist of one stage in the unfolding procedure. Notably, the initial input is defined as $\vv^{(0)}=\Phimat^{T}\yv$.

\begin{algorithm}[ht]
\caption{Optimization of GAP-SRN} \label{algo: gap-srn optimization}
\KwIn{All parameters of $S$ SRN: $\Theta$; learning rate $\eta$; Masks; Total $N$ pairs of measurements and ground truths $\{\yv_n, \fv_n\}_{n=1}^N$}
\KwOut{$\Theta^{*}$}
Obtain the sensing matrix $\Phimat$ upon masks by Eq.~\eqref{eq: sensing matrix}\;
Initialize $\vv^{(0)}_n=\Phimat^{T}\yv_n$\;
\While{not converged}
{
    \For{$s=1,...,S$}{
    Compute $\fv^{(s)}$ by Eq.~\eqref{eq: linear projection}\;
    }
    \For{$n=1,...,N$}{
    $\Theta\leftarrow\Theta-\eta\frac{\partial}{\partial \Theta}\mathcal{L}(\Theta; \fv_n, \yv_n)$\;
    }
}
\end{algorithm}

\noindent\textbf{Model Training.} The denoiser minimizes the distortion between the input and output. Thus, we have loss
\begin{equation}
\begin{aligned}
&\mathcal{L}(\Theta, \fv, \yv)=\sum\nolimits_{i=0}^{i=2}\alpha_{i}\| \fv - \vv^{(S-i)}\|_2, \\
&\textup{s.t.}~\vv^{(S-i)}=h(\theta; \fv^{(S-i-1)}),
\end{aligned}
\end{equation}
where $\fv$ is the ground truth, $h(\cdot)$ denotes arbitrary SRN with parameter $\theta_j$ and total parameters $\Theta$=$\{\theta_j\}^S_{j=1}$. We apply the same functional form as \cite{meng2020gap}, where multiple stage (\emph{i.e.}, $S$-2 $\sim$ $S$th stages) outputs are involved in the loss for a better reconstruction. The $\alpha_0$, $\alpha_1$, $\alpha_2$ are $1,0.5,0.5$, respectively. A pleasing result can be expected by training the model for sufficient iterations. We give the optimization procedure in Algorithm~\ref{algo: gap-srn optimization}.

\begin{table*}[t]
\tiny
\caption{PSNR (dB) comparison on 10 scenes in the simulation dataset. 
By default, the \texttt{CAE-SRN} represents the SRN enhanced by channel attention but without rescaling pairs, which achieves the state-of-the-art performance.}\label{Tab: psnr}
\vspace{-2.5mm}
\centering
\resizebox{\textwidth}{!}{
\setlength{\tabcolsep}{0.7mm}
\centering
\begin{tabular}{c|cccccccccc|c} 
	\toprule
	{Method} & {Scene1} & {Scene2} & {Scene3} 
	& {Scene4} & {Scene5} & {Scene6} & {Scene7} & {Scene8} & {Scene9} & {Scene10 }& {Avg.}\\
	\midrule
		U-net~\cite{ronneberger2015u}  & 28.28&24.06&26.02&36.33&25.51&27.97&21.15&26.83&26.13&25.07&26.80\\
		
		HSSP~\cite{Wang19_CVPR_HSSP}  & 31.07&26.30&29.00&38.24&27.98&29.16&24.11&27.94&29.14&26.44&28.93\\
		
		$\lambda$-net~\cite{Miao19ICCV}  & 30.82&26.30&29.42&37.37&27.84&30.69&24.20&28.86&29.32&27.66&29.25\\
		
		TSA-Net~\cite{Meng20ECCV_TSAnet}  & 31.26&26.88&30.03&39.90&28.89&31.30&25.16&29.69&30.03&28.32&30.24\\
		
		GSM~\cite{huang2021deep}  &32.38& 27.56&29.02&36.37&28.56&32.49&25.19&31.06&29.40&30.74&30.28\\
		
		PnP-DIP-HSI~\cite{meng2021self}  & 32.70&27.27&31.32&40.79&29.81&30.41&28.18&29.45&34.55&28.52&31.30\\
		
		GAP-net~\cite{meng2020gap} & 33.62&30.08&{\bf33.07}&40.94&30.77&33.60&27.41&31.25&{\bf33.56}&30.36&32.47\\
		CAE-SRN (ours) &{\bf34.05}&{\bf31.13}&32.34&{\bf41.32}&{\bf31.67}&{\bf35.36}&{\bf28.07}&{\bf33.41}&33.40&{\bf31.86}&{\bf33.26} \\
	\bottomrule
\end{tabular}}\vspace{-1.5mm}
\end{table*}

\begin{table*}[t]
\tiny
\caption{SSIM comparison on 10 scenes in the simulation datase. 
By default, the \texttt{CAE-SRN} represents the SRN enhanced by channel attention but without rescaling pairs, which achieves the state-of-the-art performance. }\label{Tab: ssim}
\vspace{-2mm}
\centering
\resizebox{\textwidth}{!}{
\setlength{\tabcolsep}{0.7mm}
\centering
\begin{tabular}{c|cccccccccc|c} 
	\toprule
	{Method} & {Scene1} & {Scene2} & {Scene3} 
	& {Scene4} & {Scene5} & {Scene6} & {Scene7} & {Scene8} & {Scene9} & {Scene10 }& {Avg.}\\
	\midrule
		U-net~\cite{ronneberger2015u}  &0.822&0.777&0.857&0.877&0.795&0.794&0.799&0.796&0.804&0.710&0.803\\
		HSSP~\cite{Wang19_CVPR_HSSP}   &0.852&0.798&0.875&0.926&0.827&0.823&0.851&0.831&0.822&0.740&0.834\\
		$\lambda$-net~\cite{Miao19ICCV}   & 0.880&0.846&0.916&0.962&0.866&0.886&0.875&0.880&0.902&0.843&0.886\\
		
		TSA-Net~\cite{Meng20ECCV_TSAnet}   & 0.887&0.855&0.921&0.964&0.878&0.895&0.887&0.887&0.903&0.848&0.893\\
		
		GSM~\cite{huang2021deep}    & 0.920&0.892&0.925&{\bf0.970}&0.894&0.938&0.898&0.932&0.925&0.934&0.923\\
		
		PnP-DIP-HSI~\cite{meng2021self}   & 0.898&0.832&0.920&{\bf0.970}&0.903&0.890&0.913&0.885&0.932&0.863&0.901\\
		
		GAP-net~\cite{meng2020gap} & {\bf0.926}&{\bf0.914}&{\bf0.944}&0.966&0.925&0.936&{\bf0.915}&0.918&{\bf0.937}&0.914&0.929\\
		CAE-SRN (ours)&0.925&0.906&0.924&0.969&{\bf0.940}&{\bf0.956}&0.871&{\bf0.952}&{\bf0.937}&{\bf0.937}&{\bf0.931} \\
	\bottomrule
\end{tabular}}
\vspace{-2mm}
\end{table*}

\begin{figure*}[t] 
\centering 
\includegraphics[width=1\textwidth]{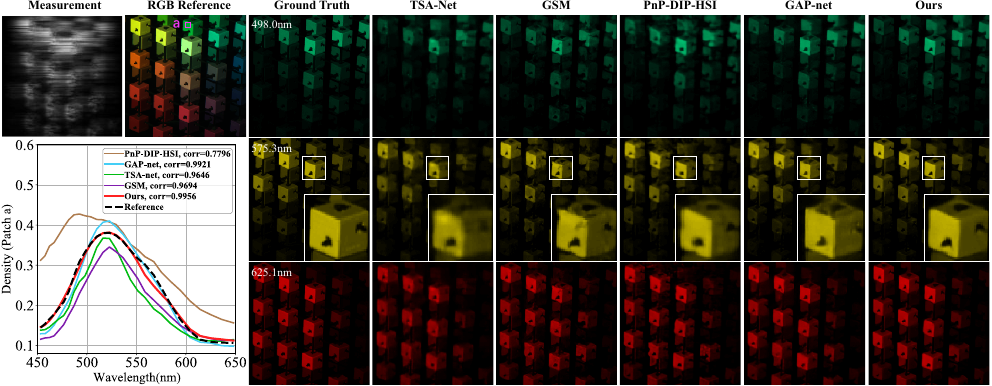}
\vspace{-4mm}
\caption{Comparison of reconstruction results for a synthetic hyperspectral image. Four state-of-the-art methods and our method (right column) are presented on 3 out of 28 spectral channels. The RGB reference is shown to demonstrate the color (top-left). The density-vs-wavelength curves (bottom-left) corresponding to the chosen patch (\emph{i.e.}, \texttt{patch a}) are plotted to demonstrate the spectral fidelity. Our results recover the most details contained in the ground truth, \emph{i.e.}, enlarged windows on 575.3nm wavelength. Please zoom in for a better visualization.} 
\label{Fig: simu1}
\vspace{-5mm}
\end{figure*}

\section{Experiments \label{sec: experiments}}
In this section, we evaluate the proposed method on both spectral and video data. We design the empirical evaluations for answering the following questions:
\begin{itemize}
    \setlength\itemsep{0em}
    \item \emph{How does the proposed CAE-SRN compare against the state-of-the-art in spectral SCI?} (Section~\ref{subsec: experiment spectral sci})
    \item \emph{Besides the reconstruction performance, what are the further advantages the proposed simple structure brings about for spectral SCI?} (Section~\ref{subsec: model discussion})
    \item \emph{How does the proposed GAP-SRN compare against the state-of-the-art in video SCI?} (Section~\ref{subsec: video reconstruction})
\end{itemize}

\begin{figure*}[t] 
\centering 
\includegraphics[width=0.97\textwidth]{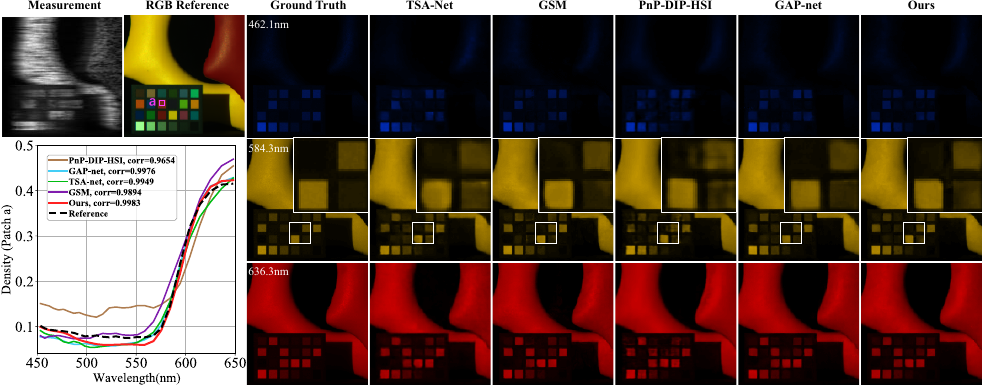}
\vspace{-2mm}
\caption{Comparison of reconstruction results for a synthetic hyperspectral image. Four state-of-the-art methods and our method (right column) are presented on 3 out of 28 spectral channels. The RGB reference is shown to demonstrate the color (top-left). The density-vs-wavelength curves (bottom-left) corresponding to the chosen patch (\emph{i.e.}, \texttt{patch a}) are plotted to demonstrate the spectral fidelity. Our results recover the most details contained in the ground truth, \emph{i.e.}, enlarged windows on 584.3nm wavelength. Please zoom in for a better visualization.} 
\label{Fig: simu2}
\vspace{-4mm}
\end{figure*}

\subsection{Experimental Settings}\label{subsec: settings}
\noindent \textbf{Spectral Dataset.} For a fair comparison, we adopt the same 28 wavelengths for the HSI as in~\cite{Meng20ECCV_TSAnet}, distributed within the range of 450nm to 650nm obtained by spectral interpolation manipulation, and conduct experiments on the following two datasets: 1) Simulation Data and 2) Real HSI data. 
For simulation, both CAVE~\cite{CAVE_spectraldata_07} synthetic dataset and KAIST~\cite{choi2017high} synthetic dataset are applied in our simulation experiment. For the training set, we create 205 1,024$\times$1,024$\times$28 large image examplers from 30 256$\times$256$\times$28 images from CAVE dataset by randomly concatenating. Operations like rotation and rescaling are both used for better generalization. During training, we randomly crop 256$\times$256$\times$28 samples from the examplers. For testing set, ten benchmark 256$\times$256$\times$28 HSIs abstracted from KAIST dataset are applied, following~\cite{Meng20ECCV_TSAnet,huang2021deep}. For real HSI reconstruction, we expand the previous training set by adding 37 HSI images from KAIST~\cite{choi2017high} dataset. The spatial size of training samples becomes 660$\times$660$\times$28 to be consistent with the real-world measurements, which are obtained by the SD CASSI system developed in~\cite{Meng20ECCV_TSAnet}.

\noindent \textbf{Video Dataset.} We adopt the DAVIS2017~\cite{pont20172017} dataset as the training set of the video SCI. The total number of 26,000 videos of the shape 256$\times$256$\times$8 are generated by data augmentation techniques, \emph{i.e.}, cropping, rotation and downsampling, following~\cite{meng2020gap}. We test the model on benchmark simulation dataset~\cite{yuan2020plug}, which consists of six videos with the same spatial size (256$\times$256) and different frame numbers, including \texttt{Kobe}, \texttt{Drop}, \texttt{Traffic}, \texttt{Aerial}, \texttt{Vehicle} and \texttt{Runner}.

\noindent \textbf{Compared Methods}. For spectral SCI, we compare with seven state-of-the-art reconstruction algorithms, including the U-Net~\cite{ronneberger2015u}, HSSP~\cite{Wang19_CVPR_HSSP}, $\lambda$-net~\cite{Miao19ICCV}, TSA-Net~\cite{Meng20ECCV_TSAnet}, GSM~\cite{huang2021deep}, PnP-DIP-HSI~\cite{meng2021self} and GAP-net~\cite{meng2020gap} among which the most recent PnP-DIP-HSI and GAP-net demonstrate best performances. For video SCI, we compare with six popular methods on aforementioned benchmark testing set, including GAP-TV~\cite{yuan2016generalized},  E2E-CNN~\cite{qiao2020deep}, PnP-FFDNet~\cite{yuan2020plug}, DeSCI~\cite{Liu18TPAMI}, BIRNAT~\cite{cheng2020birnat}, GAP-net~\cite{meng2020gap}. Notably, the best-performed GAP-net, GAP-net-Unet-S12, is employed for final performance comparison. Standard validation criterias, Peak Signal-to-Noise Racial (PSNR) and the Structural SIMilarity (SSIM)~\cite{Wang04imagequality}, are used for a quantitative comparison. The PSNR is computed by  
\begin{equation}\label{eq: psnr compute}
\textup{PSNR}_{ch} =10\log_{10}(\frac{\textup{MAX}^2_{I}}{\textup{MSE}_{ch}}),
\end{equation}
where we firstly compute the channel-wise PSNR values (\emph{i.e.}, $\textup{PSNR}_{ch}$) and then do average. For a certain channel $ch$, ${\textup{MAX}^2_{I}}$ denotes the maximum pixel value in ground truth image $I$. By default, all PSNR values in this work following this formula.

We implemented our model by PyTorch and employed the Adam optimizer ~\cite{kingma2014adam} with $\beta_1$=0.9, $\beta_2$=0.999. We put 16 Residual Blocks (\emph{i.e.}, $M$=16) in the main body. The learning rate is initialized as 4$\times10^{-4}$ and decreased by half every 50 epochs. We set batch size as 4 for the best performance. The training takes at most 16 hours with one NVIDIA TIATN RTX GPU.

\begin{figure*}[t] 
\centering 
\includegraphics[width=0.97\textwidth]{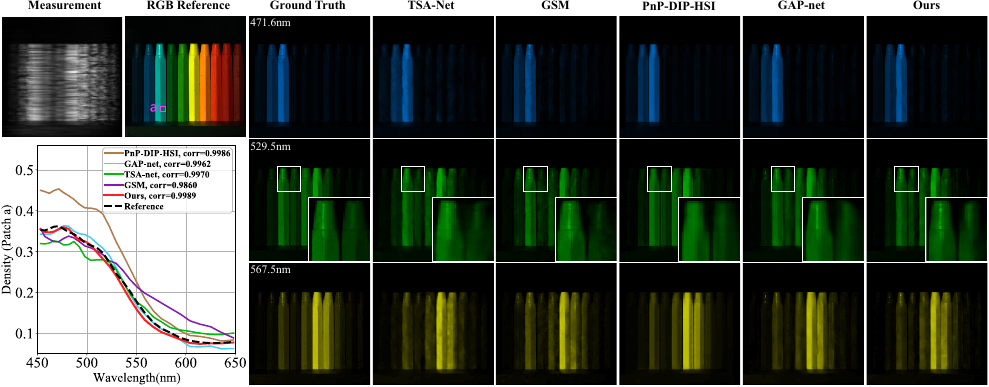}
\vspace{-2mm}
\caption{Comparison of reconstruction results for a synthetic hyperspectral image. Four state-of-the-art methods and our method (right column) are presented on 3 out of 28 spectral channels. The RGB reference is shown to demonstrate the color (top-left). The density-vs-wavelength curves (bottom-left) corresponding to the chosen patch (\emph{i.e.}, \texttt{patch a}) are plotted to demonstrate the spectral fidelity. Our results recover the most details contained in the ground truth, \emph{i.e.}, enlarged windows on 529.5nm wavelength. Please zoom in for a better visualization.} 
\label{Fig: simu3}
\vspace{-3mm}
\end{figure*}

\begin{figure*}[h] 
\centering 
\includegraphics[width=.92\textwidth]{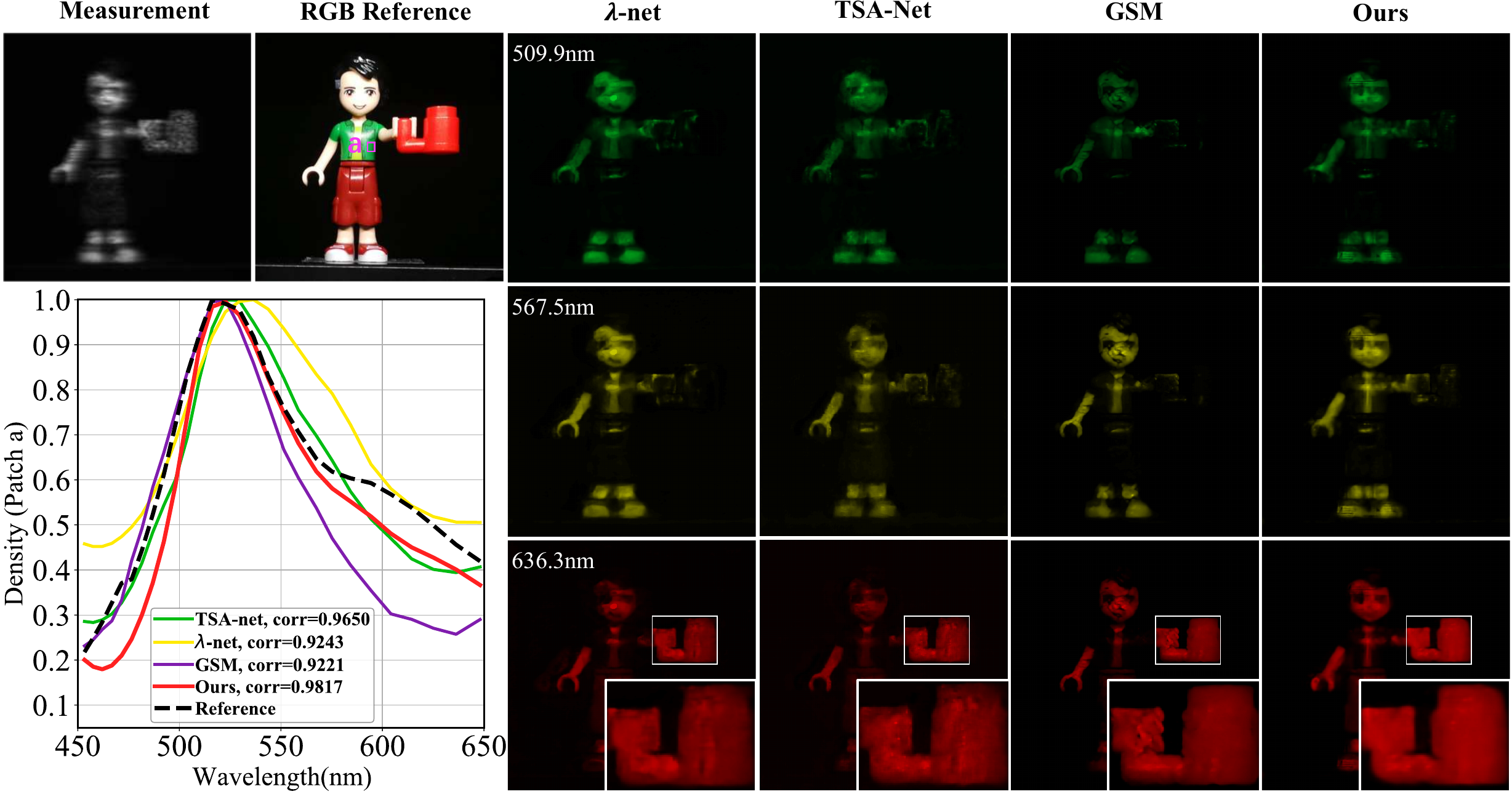}
\vspace{-2mm}
\caption{Comparison of reconstruction results for real-captured hyperspectral signal. The $\lambda$-net, TSA-Net, GSM and our method are presented on 3 out of 28 spectral channels. 
The density-vs-wavelength curves (bottom-left) corresponding to the chosen patch (\emph{i.e.}, \texttt{patch a}) are plotted to demonstrate the spectral fidelity. Notably, we employ the referenced data from a spectrometer provided by~\cite{Meng20ECCV_TSAnet}. The visual superiority of our method can be demonstrated via observing enlarged windows on 636.3nm wavelength. Please zoom in for a better visualization.} 
\label{Fig: real1}
\vspace{-5mm}
\end{figure*}

\subsection{Experiments on Sepctral SCI} \label{subsec: experiment spectral sci}
In this section, we demonstrate the performance of proposed CAE-SRN on both simulation data (Section~\ref{subsubsec: synthetic}) and real data (Section~\ref{subsubsec: real}). We apply the proposed method on the scenario of high-resolution HSI reconstruction for the first time (Section~\ref{subsubsec: high resolution}).

\begin{figure*}[ht] 
\centering 
\includegraphics[width=.94\textwidth]{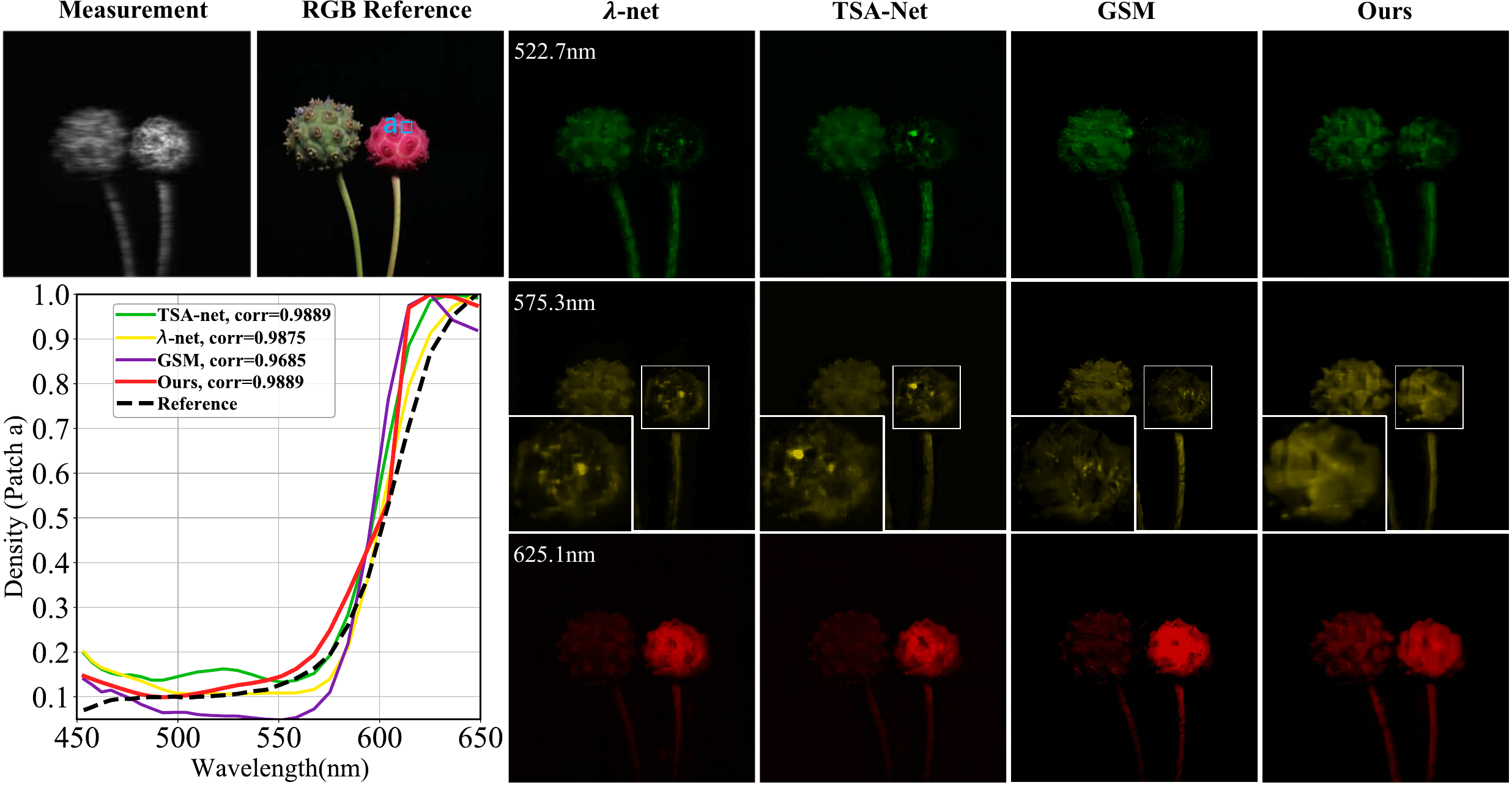}
\vspace{-1mm}
\caption{Comparison of reconstruction results for real-captured hyperspectral signal. The $\lambda$-net, TSA-Net, GSM and our method are presented on 3 out of 28 spectral channels. 
The density-vs-wavelength curves (bottom-left) corresponding to the chosen patch (\emph{i.e.}, \texttt{patch a}) are plotted to demonstrate the spectral fidelity. Notably, we employ the referenced data from a spectrometer provided by~\cite{Meng20ECCV_TSAnet}. The visual superiority of our method can be demonstrated via observing enlarged windows on 575.3nm wavelength. Please zoom in for a better visualization.} 
\label{Fig: real3}
\vspace{-5mm}
\end{figure*}

\subsubsection{HSI Reconstructions on Synthetic Data} \label{subsubsec: synthetic}

We compare with state-of-the-art methods on synthetic HSI data quantitatively and qualitatively. By default, we employ the best-performed model, \emph{i.e.}, CAE-SRN for evaluation, in which each Residual Block contains channel attention enhancement module and there are no rescaling pairs introduced. As shown in Table~\ref{Tab: psnr} and Table~\ref{Tab: ssim}, we outperform the others by a large margin. Specifically, we boost around 0.8dB/0.002 compared with the previously best method, GAP-net, in terms of PSNR/SSIM. Besides, we surpass the most recent E2E deep learning model, GSM, by nearly 3dB/0.01 regarding PSNR/SSIM but bearing much less parameters and computations (please refer to Section~\ref{subsec: model discussion} for model size comparison). Notably, we evaluate the GSM upon reconstruction results provided by the authors using Eq.~\ref{eq: psnr compute}, which leads to updated PSNR values. Overall, the comparison indicates that solely applying the proposed reconstruction backbone already enables a quantitatively well performance.

We further visualize the reconstruction results by 4 out of 7 best methods and ours in Fig.~\ref{Fig: simu1}$\sim$\ref{Fig: simu3}. For each visualization, we randomly choose 3 out of 28 spectral channels due to the limited space. Compared with the ground truth, we perceptually retrieve the most textures but introducing the least artifacts and distortion. This can be easily verified by checking enlarged windows at certain spectral channels. As for the other methods, TSA-Net avoids producing artifacts but tends to blur the scene. The GSM is good at reconstructing edges but may generate extra cracks. The PnP-DIP-HSI works globally well but shows instability among different regions. We also evaluate the spectra fidelity by plotting and comparing density-vs-wavelength curves. In a chosen region, the curves by the proposed method share the most overlaps and similar tendencies with the reference, yielding highest correlation values. Notably, we randomly choose the small patches containing varies colors for plotting, demonstrating the wavelength robustness of the proposed method. All in all, our model clearly exceeds existing methods regarding simulation hyperspectral data. In the next section, we validate the effectiveness of the proposed model on HSI data corresponding to real-world scenes collected by SD CASSI.

\begin{figure*}[ht] 
\centering 
\includegraphics[width=0.98\linewidth]{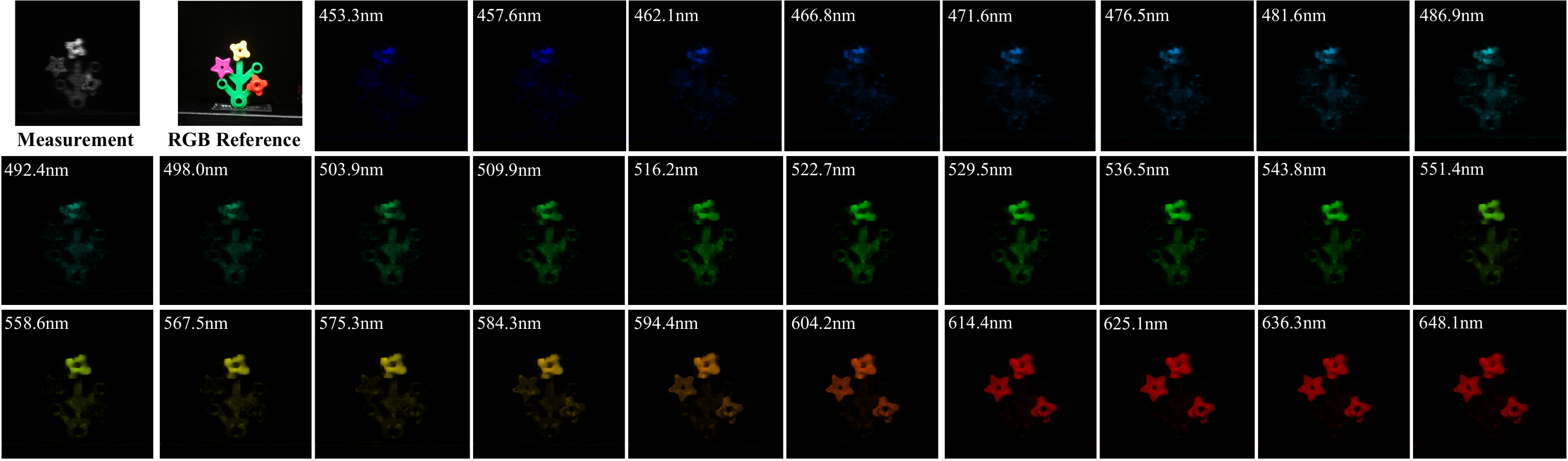}
\vspace{-2mm}
\caption{Reconstruction results for a real-captured hyperspectral signal by the proposed method. Total 28 wavelengths are included. Different channels highlight regions with distinct colors in the RGB reference (top-left).} 
\label{Fig: real2}
\vspace{-1.5mm}
\end{figure*}

\subsubsection{HSI Reconstructions on Real Data} \label{subsubsec: real} 
We still take the CAE-SRN for real HSI reconstruction. No rescaling pairs are introduced. Unlike the simulation case, it is hard to reproducing plausibly well reconstruction results by other methods through training from scratch. Therefore, we choose the methods with pre-trained models available, resulting in $\lambda$-net, TSA-Net and GSM. We perceptually compare different methods by two scenes shown in Fig.~\ref{Fig: real1} and Fig.~\ref{Fig: real3}. A quantitative comparison fails as the ground truth real HSI data is inaccessible. Similar to simulation case, we plot the density curves for chosen patches. Note that the referenced curve is computed based on the values detected by a real spectrometer following~\cite{Meng20ECCV_TSAnet,meng2021self}.

\begin{figure*}[ht] 
\centering 
\includegraphics[width=1\linewidth]{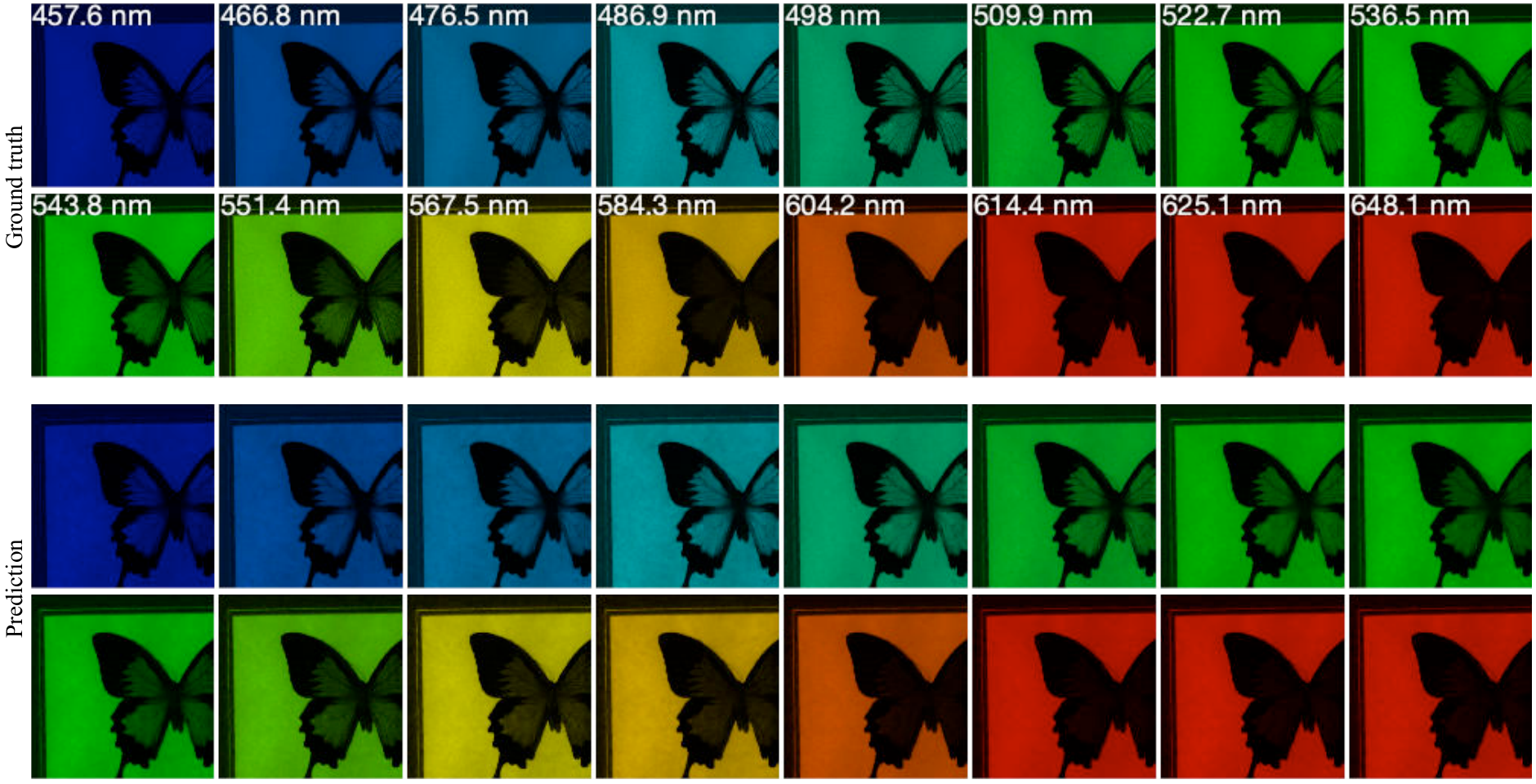}
\vspace{-5mm}
\caption{Reconstruction of the 1,024$\times$1,024 synthetic spectral data by the proposed method. Compared with the ground truth, our method enables fine-grained contents retrieval among diverse wavelengths. Zoom in for a better visualization. To our best knowledge, this is the first exhibition of the reconstruction over such a spatial scale.} 
\label{Fig: spectral_hd}
\vspace{-4.5mm}
\end{figure*}

Real HSI reconstruction is considered to be challenger than simulation case. The real captured measurements unavoidably contain noise and the images are generally characterized by more visual details in a specific spatial size (\emph{i.e.}, in our case, 660$\times$660). Bearing these difficulties, existing methods present promising results but with their own drawbacks, respectively. For example, the $\lambda$-net contains least details. Some mottled regions could be observed (\emph{i.e.}, stem regions of the plants on 522.7nm wavelength in Fig.~\ref{Fig: real3}) in reconstructions by TSA-Net. GSM avoids above problems but may cause overexposure/underexposure problems (\emph{i.e.}, ball regions by GSM in Fig.~\ref{Fig: real3}). By comparison, our method comprehensively reconstructs the original scene with less artifacts, more details and proper brightness (\emph{i.e.}, enlarged windows in Fig.~\ref{Fig: real1} and Fig.~\ref{Fig: real3}). As for the density curves, note that all methods more or less deviate from the referenced curves, among which our approximated curves are most correlated by shapes and tendencies. In Fig.~\ref{Fig: real2}, we provide complete reconstruction results on real-captured scene \texttt{lego flower}. As can be seen, differently colored regions are emphasized by distinct wavelengths accordingly, indicating an effective spectral response of proposed method. To sum up, considering the substantial dataset shift between the synthetic training set and the real testing set, our model possesses a better generalization ability, supported by aforementioned superior performance.

\begin{table}[t]
\normalsize
\caption{Model size, computation and performance comparison. The spatial size of the input is 256$\times$256. The annotation \texttt{v1}: w/o rescaling pair, \texttt{v2}: with 1 rescaling pair, \texttt{v3}: with 2 rescaling pairs and \texttt{CAE}: channel attention enhancement. All the other settings are kept the same for a fair comparison. }\label{Tab: model size}
\vspace{-4mm}
\begin{center}
\resizebox{.46\textwidth}{!}{
\begin{tabular}{lrrr}
\toprule
Method & \#params (M) & FLOPs (G) & PSNR (dB)\\
\midrule
U-net~\cite{ronneberger2015u}  & 31.32 & 58.99 & 26.80\\
$\lambda$-net~\cite{Miao19ICCV} & 62.64 & 117.98 & 29.25\\
TSA-Net~\cite{Meng20ECCV_TSAnet}  & 44.25 & 80.08 & 30.24\\
GSM~\cite{huang2021deep} & 3.76 & 646.35 & 30.28\\
\midrule
SRN \texttt{v1} & \textbf{1.25} & 81.84 & \textbf{33.17}\\
SRN \texttt{v2} & 1.44 & 25.07 & 31.77\\
SRN \texttt{v3} & 1.62 & \textbf{18.57} & 31.36\\
\midrule
CAE-SRN \texttt{v1} & \textbf{1.31} & 81.84 & \textbf{33.26}\\
CAE-SRN \texttt{v2} & 1.49 & 25.07 & 32.33\\
CAE-SRN \texttt{v3} & 1.68 & \textbf{18.57} & 32.07\\
\bottomrule
\end{tabular}}
\end{center}
\vspace{-5mm}
\end{table}

\begin{table}[t]
\tiny
\caption{Discussion on number of Residual Blocks. All the other settings are kept the same for a fair comparison. No rescaling pairs are involved in the model.}\label{Tab: RB number}
\vspace{-5mm}
\begin{center}
\resizebox{.38\textwidth}{!}{
\begin{tabular}{lrr}
\toprule
\#Residual Blocks & PSNR (dB) & SSIM\\
\midrule
8 &32.02&0.908\\
12 &32.75&0.924\\
16 &33.17&0.929\\
\bottomrule
\end{tabular}}
\end{center}
\vspace{-5mm}
\end{table}

\begin{table}[t]
\caption{Robustness to real masks by different methods. The settings include \texttt{one-to-one}: training/testing model upon identical mask, \texttt{one-to-many}: models trained by the same mask are tested by diverse masks and \texttt{many-to-many}: training/testing the model with random masks. We report (averaged) PSNR (dB) and perform 100 trials for random testing. }\label{Tab: robustness to mask}
\vspace{-5mm}
\begin{center}
\resizebox{.46\textwidth}{!}{
\begin{tabular}{lccc}
\toprule
Method & one-to-one & one-to-many & many-to-many\\
\midrule
TSA-Net & 30.24 & 27.47$\pm$0.46 & 21.42$\pm$0.07\\
GSM     & 30.28 & 26.34$\pm$0.06 & 28.20$\pm$0.01\\
Ours    & 33.17 & 30.17$\pm$0.63 & 32.24$\pm$0.10\\
\bottomrule
\end{tabular}}
\end{center}
\vspace{-8mm}
\end{table}

\subsubsection{High-resolution HSI Reconstruction} \label{subsubsec: high resolution} 
\begin{figure}[t]
\begin{center}
\includegraphics[width=0.98\linewidth]{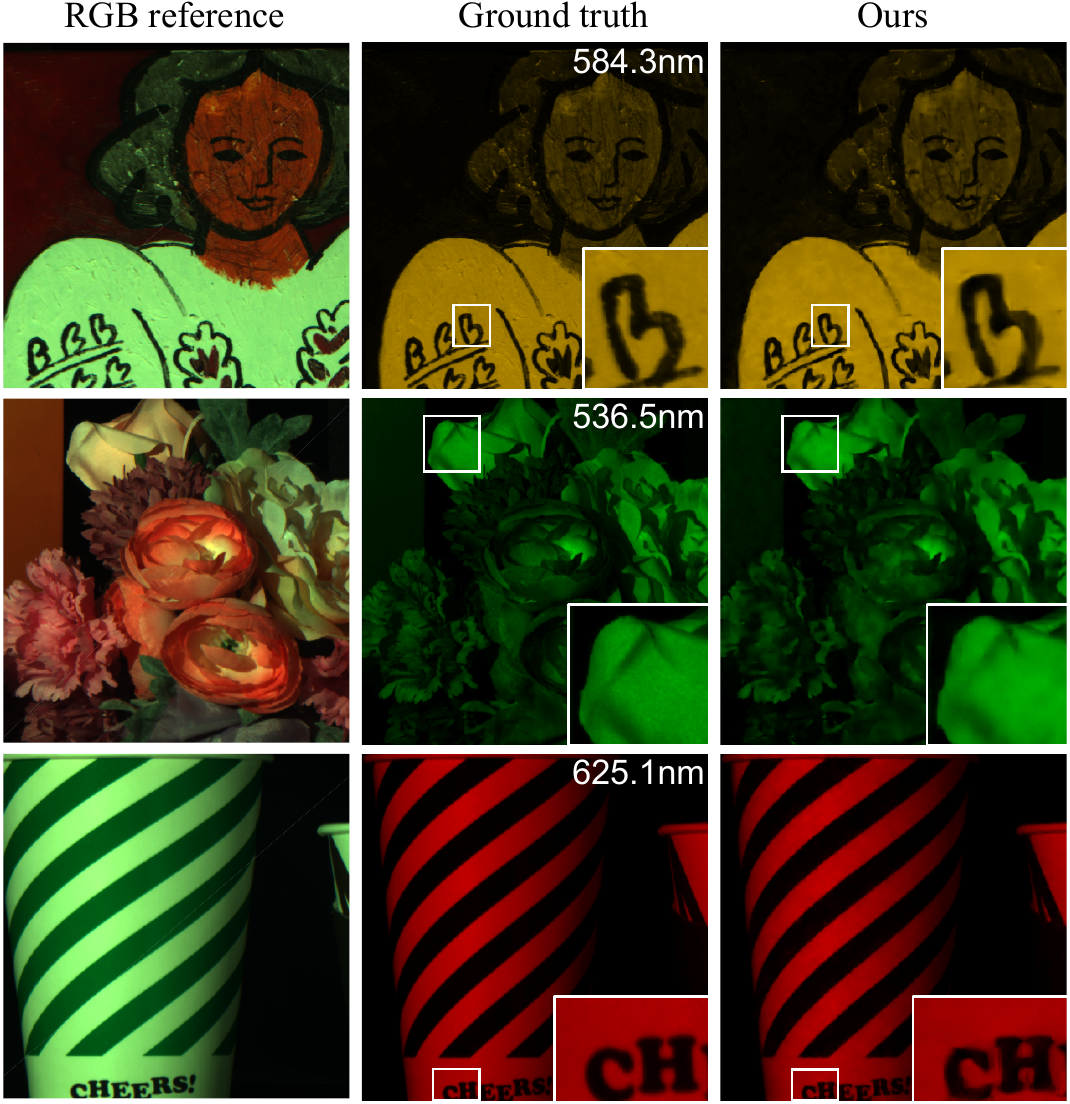}\vspace{-5mm}
\end{center}
\caption{Visualization of 1,024$\times$1,024 HSI reconstruction. By enlarged windows, the proposed method (right column) could produce plausible visual details compared with the ground truth (middle column). RGB reference (left column) are shown to present the color.}
\label{fig: details_hr_reconstruction}
\vspace{-6mm}
\end{figure}

In this section, we perform high-resolution HSI reconstruction (\emph{i.e.}, 1,024$\times$1,024$\times$28) upon SRN. We build a synthetic high-resolution HSI testing set from the KAIST dataset~\cite{choi2017high}. Technically, existing methods are all capable of reconstructing HSI with arbitrary spatial size. However, it is practically difficult due to the huge requirement of the computation resource. For example, given an input of 256$\times$256$\times$28, the TSA-Net containing 44.25 million parameters needs over 8$\times 10^{10}$ computations (see detailed discussion on computation in Section~\ref{subsec: model discussion}), leading to 18.7Gb GPU memory take-up. For an input of 1,024$\times$1,024$\times$28, the model size remain unchanged while the computation amount would exceed 1$\times 10^{12}$. Even two 24Gb GPUs fails to support such a computation. By comparison, the SRN with smaller computation burden makes high-resolution HSI reconstruction more feasible. We compare the reconstruction with the ground truth upon 16 out of 28 spectral channels in Fig.~\ref{Fig: spectral_hd}. The proposed backbone continuously produces plausible results under a large spatial size. In Fig.~\ref{fig: details_hr_reconstruction}, we visually present more reconstruction results on random spectral channels. We globally produce plausible results with fine-grained textures as the ground truth. As compared by enlarged windows, the difference between ours and the ground truths are imperceptible, showcasing the strong reconstruction capacity of the proposed backbone toward high-resolution HSIs. To our best knowledge, this is the first display of reconstructed spectral data with over 1000-pixel spatial scale and a large spectral channel number.

\subsection{Model Discussion} \label{subsec: model discussion}
In this section, we analysis key components upon simulated spectral data by ablation studies and uncover some advantages attributing to the simple network.

\begin{table*}[t]
\normalsize
\caption{PSNR (dB) and SSIM values by different algorithms on 6 benchmark simulation video data. The proposed SRN is employed as a deep reconstruction backbone for the unfolding framework GAP, denoted by \texttt{GAP-SRN}.
}\label{Tab: video performance}
\vspace{-2.5mm}
\centering
\resizebox{\textwidth}{!}{
\setlength{\tabcolsep}{0.7mm}
\centering
\begin{tabular}{l|cccccccccccc|cc} 
	\toprule
	\multirow{2}{*}{Method} & \multicolumn{2}{c}{Kobe} & \multicolumn{2}{c}{Traffic} & \multicolumn{2}{c}{Runner} 
	& \multicolumn{2}{c}{Drop} & \multicolumn{2}{c}{Vehicle} & \multicolumn{2}{c}{Aerial} &\multicolumn{2}{|c}{Average}\\
	\cmidrule(lr){2-3}
	\cmidrule(lr){4-5}
	\cmidrule(lr){6-7}
	\cmidrule(lr){8-9}
	\cmidrule(lr){10-11}
	\cmidrule(lr){12-13}
	\cmidrule(lr){14-15}
	&PSNR&SSIM&PSNR&SSIM&PSNR&SSIM&PSNR&SSIM&PSNR&SSIM&PSNR&SSIM&PSNR&SSIM\\
	\midrule
		GAP-TV~\cite{yuan2016generalized} &26.46&0.885&20.89&0.715&28.52&0.909&34.63&0.970&24.82&0.838&25.05&0.828&26.73&0.858\\
		E2E-CNN~\cite{qiao2020deep} &29.02&0.861&23.45&0.838&34.43&0.958&36.77&0.974&26.40&0.886&27.52&0.882&29.26&0.900\\
		PnP-FFDNet~\cite{yuan2020plug} &30.50&0.926&24.18&0.828&32.15&0.933&40.70&0.989&25.42&0.849&25.27&0.829&29.70&0.892\\
		DeSCI~\cite{Liu18TPAMI} &{\bf33.25}&0.952&28.72&0.925&{\bf38.76}&0.969&43.22&0.993&25.33&0.860&27.04&0.909&32.72&0.935\\
		GAP-net-Unet-S12~\cite{meng2020gap}&32.09&0.944&28.19&0.929&38.12&0.975&42.02&0.992&27.83&0.931&28.88&0.914&32.86&0.947\\
		BIRNAT~\cite{cheng2020birnat} &32.71&0.950&29.33&0.942&38.70&0.976&42.28&0.992&27.84&0.927&28.99&0.917&33.31&0.951\\
		GAP-SRN (ours) &32.79&{\bf0.954}&{\bf29.37}&{\bf0.946}&38.66&{\bf0.976}&{\bf43.35}&{\bf0.999}&{\bf28.09}&{\bf0.940}&{\bf29.31}&{\bf0.924}&{\bf33.59}&{\bf0.959}\\
	\bottomrule
\end{tabular}}\vspace{-5mm}
\end{table*}

\begin{figure*}[ht] 
\centering 
\includegraphics[width=0.98\linewidth]{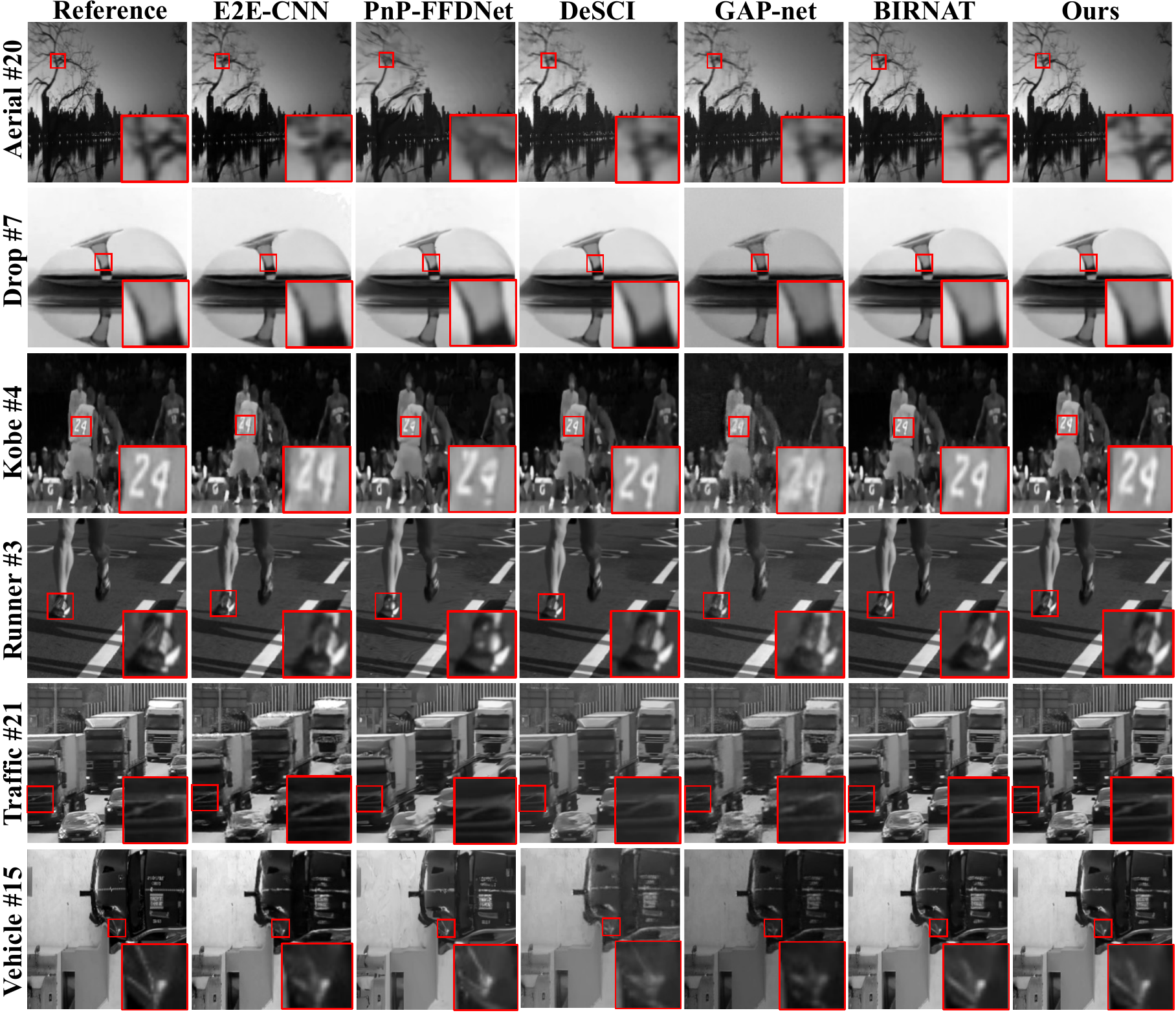}
\vspace{-2mm}
\caption{Reconstruction of the synthetic video data. Methods are demonstrated by random chosen frames. Compared with the ground truth (left column), both BIRNAT~\cite{cheng2020birnat} and our method recover the most textural details.} 
\label{Fig: video_simu}
\vspace{-5mm}
\end{figure*}

\noindent \textbf{Ablation Study}. We perform two ablation studies to verify the effectiveness of the key components. Firstly, we focus on the validity of the channel attention module by comparing between SRN and CAE-SRN among three versions, \emph{i.e.}, \texttt{v1}: both introduce no rescaling pair, \texttt{v2}: both equally contain one rescaling pair, \texttt{v3}: both equally apply two rescaling pairs. As shown in Table~\ref{Tab: model size}, the channel attention enhancement (CAE) module facilities evident performance boosts over the latter two versions, \emph{i.e.}, 0.56dB/0.71dB by comparing with the single SRN over \texttt{v2}/\texttt{v3} models. Notably, only a slight performance boost (0.09dB) can be observed by comparing with two \texttt{v1} models. Therefore, one can derive a more intuitive conclusion by jointly considering three versions -- the CAE module serves as a remedy for performance descent caused by rescaling pairs. This is consistent with the discovery revealed by~\cite{zhang2021accurate} as we previously mentioned in Section~\ref{subsec: CAE+SRN}. In conclusion, the CAE module contributes to the final performance, especially for the cases with rescaling pairs.

The other fundamental component of the proposed method is the Residual Block (RB). One can simply verify the effectiveness by controlling the number of RBs in SRN network. We exploit three cases including 8, 12 and 16. As reported in Table \ref{Tab: RB number}, better performance can be expected with more RBs as there are more parameters included and a larger network depth obtained. Note that the smallest model with 8 RBs achieves over 32dB regarding PSNR, supposing a promising performance of the proposed structure as a lightweight backbone.

\noindent \textbf{Model Size and FLOPs}. As compared in Table~\ref{Tab: psnr}, seminal works achieve promising performance in HSI reconstruction. However, one of their problem is the large model size, undermining further deployments to platforms with limited memories. Besides, the giant model volume obstacles the potential incorporation with novel techniques or frameworks, restricting the further development of these insightful works. One possible solution of reducing the model size while maintaining the performance is to employ model compression techniques~\cite{cheng2017survey}, while operating on the highly specialized structures remains to be quite challenge. By analysis, a compact reconstruction model with a competitive performance is required and our proposed backbone, SRN, properly meets the requirement. In Table~\ref{Tab: model size}, we compare the model size (number of trainable parameters) of popular deep learning-based methods and variants of SRN. Our best model, \texttt{CAE-SRN v1}, only takes no more than 35\% parameters of GSM while boosts nearly 3dB as for PSNR. By comparing with TSA-Net, we only take less than 3\% parameters while enjoy a more significant improvement. Our largest model, \texttt{CAE-SRN v3} contains additional 0.43 million parameters due to the channel attention enhancement module and \texttt{CONV} layers in rescaling pairs, which turns out to be negligible.

For different methods, we also measure the Floating-Point Operations (FLOPs). Generally, large model size leads to more computations, \emph{i.e.}, the amount of additions and multiplications, thus resulting in high FLOPs. In addition, the spatial size of the input also impacts the FLOPs. Recall that in Section~\ref{subsubsec: high resolution} we utilize this property. We report the FLOPs of different methods in Table~\ref{Tab: model size}, treating it as a reference for measuring time efficiency. Our best performed model, \texttt{CAE-SRN v1}, only takes 12.7\% FLOPs of GSM. By introducing rescaling pairs in \texttt{CAE-SRN v3}, we achieve 2.9\% FLOPs of GSM while obtain over 32dB performance. Besides, the rescaling pairs notably reduce the FLOPs while the CAE module barely induces the computation growth. In summary, carrying small model sizes and extremely low FLOPs, the proposed backbone not only demands an economical memory usage, but also is potentially applicable to latency-demanding scenarios.

\noindent \textbf{Robustness on Masks.} According to Eq.~\eqref{eq: vec model}, the forward model of SCI is determined by the sensing matrix based on the real mask. Although previous methods offer satisfying approximations to the inverse model, they suffer from performance degradation when applied to new forward models (optical systems with unseen masks), showcasing constrained robustness on masks. The underlying interpretation is the overconfidence owning to their strong modeling capacities. In contrast, simpler model is less troubled~\cite{guo2017calibration}. We compare the robustness on real masks of different methods in Tab.~\ref{Tab: robustness to mask}. Specifically, for single mask-trained models, we test with random real masks, namely \texttt{one-to-many}. To make a further step, we train and test models with random masks, namely \texttt{many-to-many}. We name the trainditional setting as \texttt{one-to-one}. Note that masks for training and testing are disjoint. By conclusion, our model indeed enjoys higher robustness over TSA-Net and GSM, regardless of the training setting on masks. Specifically, training the model with random masks leads to over 2dB improvement, enabling a flexible adaptation of the proposed backbone to multiple optical systems.

\subsection{Experiments on Video SCI}\label{subsec: video reconstruction}
In this section, we perform experiments on video SCI by integrating the SRN backbone into the GAP framework. The performance of the GAP-SRN is analyzed. Similar to spectral SCI, we also uncover the superiority of GAP-SRN on model size and computation burden.

\begin{figure}[t]
\begin{center}
\includegraphics[width=0.99\linewidth]{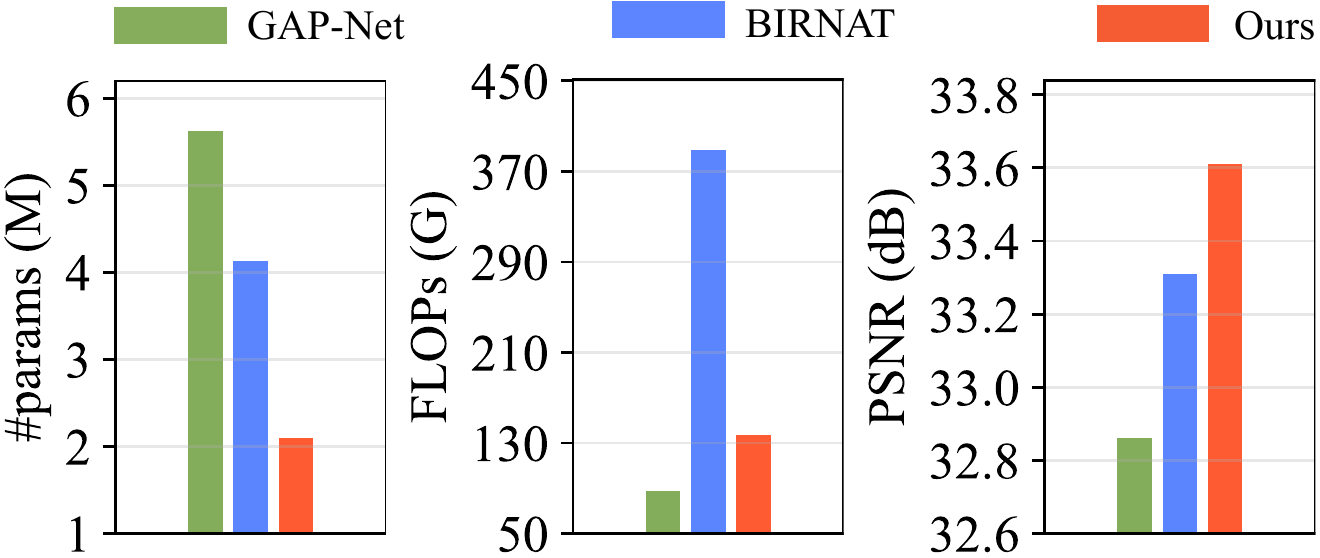}\vspace{-5mm}
\end{center}
\caption{Comparison between different methods for the video SCI. The model size (\textit{left}) is shown by the amount of parameters ($\downarrow$), the computation burden (\textit{middle}) is measured by FLOPs ($\downarrow$) and the performance (\textit{right}) is indicated by PSNR ($\uparrow$). The proposed method performs best with a smallest model size.}
\label{fig: video_compare}
\vspace{-8mm}
\end{figure}

\noindent\textbf{Performance Comparison}. In Table~\ref{Tab: video performance}, we report the performance of existing famous methods. By default, we use the SRN containing 16 Residual Blocks and best-performed GAP-net-Unet-S12 (simplified as GAP-Net) throughout the experiment. No rescaling pairs are introduced. We rank first on 4 out of 6 benchmark video data. The SRN is superior than backbone U-Net, which is evidenced by the performance boost over GAP-Net (\emph{i.e.}, 0.73dB/0.012 in terms of PSNR/SSIM). Notably, all the other settings are kept the same for a fair comparison. Besides, we also visualize the reconstructed video frames in Fig.~\ref{Fig: video_simu}, where we exhibit random frames chosen from the video data (rows) and compare six methods (columns). Our method produces more textures and proper contrast as indicated by the enlarged windows. Note that the BIRNAT shares little visual difference as ours, which is consistent with the little performance gap between both.

\noindent \textbf{Model Discussion}. The outstanding performance of GAP-SRN is gained at no cost of the model size. In Fig.~\ref{fig: video_compare}, we further compare the model size (by the number of parameters), computation burden (by FLOPs) and the performance (by PSNR) of proposed GAP-SRN, GAP-Net and the BIRNAT. For the model size, we only uses half amount of parameters contained in the BIRNAT (\emph{i.e.}, 4.13M v.s. 2.10M). In terms of the computation, both our method and the GAP-net are within the same scale (\emph{i.e.}, around or lower than 130G), while BIRNAT takes 3 times as ours. Notably, we achieve 0.75dB performance improvement by exchanging the backbone in original GAP-net, also exceed BIRNAT by 0.3dB in PSNR. Taken together, we surpass the compared methods in terms of the triple trade-off among model size, computation burden and the performance.

\section{Conclusion \label{sec: conclusion}}
We have provided a simple yet highly-efficient backbone for the SCI community. The proposed backbone filled the research gap of residual learning-based model design. For the application of HSI, we retrieved the missing visual informative details via exploiting more clues provided by the spectral coherence, leading to CAE-SRN. For video reconstruction, we combined the backbone with a popular unfolding framework, generalized alternating projection (GAP), to produce high-fidelity results and meanwhile, minimize the inference time. On both applications, the leading performances demonstrated the efficiency of the proposed backbone, and the corresponding versatility. We hope the proposed network design with the underlying insights will benefit the future work in the emerging study of SCI.

\bibliographystyle{splncs}
\bibliography{main}   

\end{document}